\def\year{2022}\relax
\documentclass[letterpaper]{article} 
\usepackage{aaai22}  
\usepackage{times}  
\usepackage{helvet}  
\usepackage{courier}  
\usepackage[hyphens]{url}  
\usepackage{graphicx} 
\usepackage{natbib}  
\usepackage{caption} 
\DeclareCaptionStyle{ruled}{labelfont=normalfont,labelsep=colon,strut=off} 
\usepackage{subfigure}
\usepackage{enumitem}
\usepackage{comment}
\usepackage{array}
\usepackage{mathtools}
\usepackage{amsmath}
\usepackage{amssymb}
\usepackage{color}
\usepackage{multirow}
\usepackage{balance}
\usepackage{threeparttable}
\usepackage{booktabs} 
\usepackage{epsfig}
\usepackage{arydshln}
\usepackage{url}
\usepackage{bbold}\usepackage{diagbox}
\input{insbox}

\usepackage{epstopdf}
\makeatletter
\newif\if@restonecol
\makeatother

\usepackage[linesnumbered, ruled, vlined]{algorithm2e}

\newcommand{\hide}[1]{} 
\newcommand{\vpara}[1]{\vspace{0.05in}\noindent \textbf{#1 }}
\newcommand{\ipara}[1]{\vspace{0.03in}\noindent \textit{#1 }}

\newcommand{\figref}[1]{Figure~\ref{#1}} 
\newcommand{\beq}[1]{\vspace{-0.03in}\begin{equation}#1\end{equation}\vspace{-0.03in}}
\newcommand{\beqn}[1]{\vspace{-0.04in}\begin{eqnarray}#1\end{eqnarray}\vspace{-0.04in}}

\newtheorem{problem}{Problem}
\newtheorem{definition}{Definition}

\newcommand{\pretrain}{CODE-pre}
\newcommand{\spretrain}{CODE-pre\space}

\newcommand{\model}{CODE}
\newcommand{\smodel}{CODE\space}

\usepackage{newfloat}
\usepackage{listings}
\title{\model: Contrastive Pre-training with Adversarial Fine-tuning for Zero-shot Expert Linking}
\author{
	Bo Chen\textsuperscript{\rm 1},
	Jing Zhang\textsuperscript{\rm 2}\thanks{Jing Zhang is the Corresponding Author.},
	Xiaokang Zhang\textsuperscript{\rm 2},
	Xiaobin Tang\textsuperscript{\rm 2},
	Lingfan Cai\textsuperscript{\rm 2},\\
	Cuiping Li\textsuperscript{\rm 2},
	Hong Chen\textsuperscript{\rm 2},
	Peng Zhang\textsuperscript{\rm 3},
	Jie Tang\textsuperscript{\rm 1}
}
\affiliations{
	\textsuperscript{\rm 1}Department of Computer Science and Technology, Tsinghua University, Beijing, China\\
    \textsuperscript{\rm 2}Information School, Renmin University of China, Beijing, China\\
    \textsuperscript{\rm 3}Zhipu.AI, Beijing, China
	
}

\usepackage{bibentry}

\begin{document}
	\maketitle

	\begin{abstract}
	Expert finding, a popular service provided by many online websites such as Expertise Finder, LinkedIn, and  AMiner, is beneficial to seeking candidate qualifications, consultants, and collaborators. However, its quality is suffered from lack of ample sources of expert information. This paper employs AMiner as the basis with an aim at linking any external experts to the counterparts on AMiner. 
	As it is infeasible to acquire sufficient linkages from arbitrary external sources, we explore the problem of zero-shot expert linking.
	In this paper, we propose \model, which first pre-trains an expert linking model by contrastive learning on AMiner such that it can capture the representation and matching patterns of experts without supervised signals, then it is fine-tuned between AMiner and external sources to enhance the model's transferability in an adversarial manner. For evaluation, we first design two intrinsic tasks, author identification and paper clustering, to validate the representation and matching capability endowed by contrastive learning. Then the final external expert linking performance on two genres of external sources also implies the superiority of the adversarial fine-tuning method. Additionally, we show the online deployment of \model, and continuously improve its online performance via active learning.

	\end{abstract}

	\section{Introduction}
\label{sec:intro}
Online  websites such as 
Expertise Finder\footnote{https://expertisefinder.com/},
LinkedIn\footnote{http://www.linkedin.com}, and  AMiner\footnote{https://www.aminer.cn/} provide valuable services of expert finding for governments or research groups to find consultants, collaborators, candidate qualifications, etc.
However, expert information is dispersed across heterogeneous sources. For example, Google Scholar and AMiner maintain academic information, LinkedIn keeps skills and background, and news articles report the real-time activities of experts. A single source of information is far from comprehensive and convincing to support the high-quality expert finding, which demands for integrating heterogeneous expert information together.
\begin{figure*}[t]
	\centering
	\includegraphics[width=1.0\textwidth]{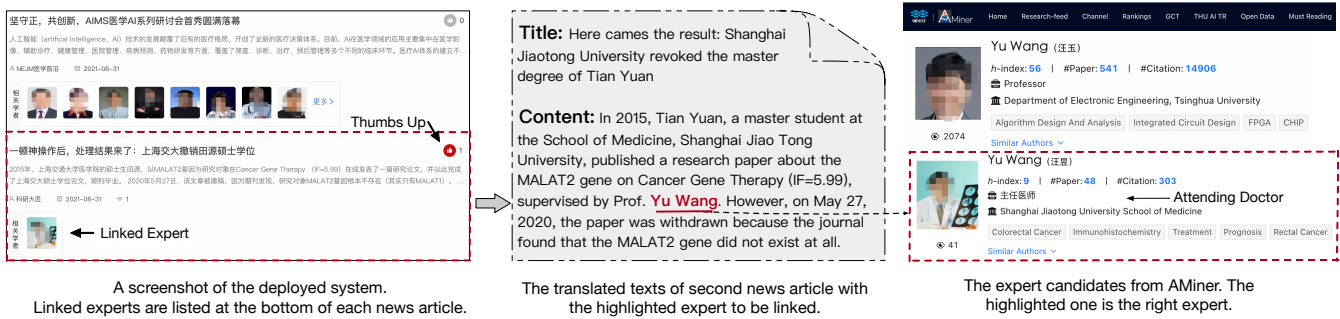}
	\caption{\label{fig:motivation} The deployed online expert linking system (Left), which aims at linking expert from news articles to AMiner. The case of linking expert ``Yu Wang" from the highlighted article is presented (Right).}
\end{figure*}
This paper employs AMiner, a free online academic search and mining system collecting over 100 million researcher profiles with 200 million papers from multiple databases~\cite{tang2008arnetminer}, as the implementation basis.
We target at linking heterogeneous expert information from external sources to AMiner. Figure~\ref{fig:motivation} illustrates the online system\footnote{https://aminer.top3-talent.com/homepage} deployed with the proposed \smodel to link an expert from a news article to the right candidate on AMiner. 


Entity linking is a related research field that links entities extracted from unstructured texts to those in a knowledge graph~\cite{ klie2020zero, hou2020improving, angell2021clustering}.
However, prevailing methods often resort to the huge amount of labeled data to encode entities from heterogeneous sources into a unified space. 
Unfortunately, the linkages between external
information and the AMiner experts are often arduous to
obtain. 
For example, in AMiner, it usually spends up to several hours to correct the collected papers for a top expert by a skilled annotator. Moreover, the external information about experts come from arbitrary sources, making it unforeseeable for us to annotate the corresponding labels beforehand.
In view of this, we pay attention to the problem of zero-shot expert linking. A natural question arises: \textit{can we learn a universal 
expert linking model from abundant AMiner experts such that it can be transferable to unseen external  expert linking?}

Besides the label scarcity issue, we also face challenges about how to represent an expert and match two experts, because:
(1) an expert, consisting of different types of information such as demographic attributes, papers, or news, is neither a continuous signal as an image nor a discrete signal as a word, which demands a non-trivial method for expert representation beyond the standard image or word representation methods; 
(2) the gap of morphology, syntax, topics between AMiner and external sources is obvious, which motivates us to fine-tune the basic expert representation model on external sources to improve its transferability. 

\vpara{Present Work.} We propose \model, a \underline{CO}ntrastive Pre-training with A\underline{D}versarial Fine-tuning for Z\underline{E}ro-shot expert Linking model, to link experts from external sources to AMiner in the zero-shot setting. To address the label scarcity issue, \smodel is first pre-trained on AMiner via contrastive learning~\cite{wu2018unsupervised}. To enable this, we define the pre-training task as expert discrimination which samples instances from each AMiner expert, and pulls the instances sampled from the same expert close together but pushes those sampled from different experts far away from each other. An expert instance is defined as a set of papers owned by the expert, which is then encoded by BERT~\cite{devlin2018bert} and matched with each other by an interaction-based fine-grained metric. 
After pre-training \model, we apply the adversarial method to adapt it to unseen external sources when linking experts from them to the AMiner experts.


We demonstrate the validity of our model by  a series of experiments, demonstrating that: (1) 
\smodel is able to represent an expert well, based on which the authors of a paper can be correctly identified (the evaluation task of author identification) and the papers belonging to the same author can be correctly clustered (the evaluation task of paper clustering); (2) \smodel is able to be transferred to unseen external sources such as news articles or LinkedIn pages to link the mentioned experts to the AMiner experts (the evaluation task of external expert linking).

We summarize our contributions as follows:
\begin{itemize}[leftmargin=*]
	
	\item 
	We propose \smodel consisting of a contrastive pre-training module and an adversarial fine-tuning module, to address zero-shot expert linking from external sources to AMiner.
	
	\item 
	We define expert discrimination as the pre-training task along with an interaction-based metric function to characterize both the universal representation and fine-grained matching patterns of experts. Adversarial learning is used to improve the transferability of the pre-trained model.

	\item In addition to the extensive experiments which demonstrate the superiority of  \model, we also deploy  \smodel online, and further involve human in the loop to improve online performance via active learning. 
	All codes and data are available
	at https://github.com/allanchen95/CODE.

\end{itemize}

\hide{We evaluate COAD by two widely studied downstream task --- author identification and paper clustering. The results show that the proposed pre-training method can surpass various baselines without contrastive learning over experts by 12.7-25.8\% in terms of HitRatio@1. Furthermore, we evaluate the complete COAD by linking two different genres of unseen external sources to AMiner experts. COAD outperforms the state-of-the-art domain adaption methods by 0.6-2.3\% in terms of HitRatio@1, which demonstrates the prominent transferability of the proposed adversarial fine-tuning method. All codes and data are publicly available\footnote{https://github.com/BoChen-Daniel/Expert-Linking}}
	\section{Related Work}
\label{sec:related}


\vpara{Entity Linking.}
Most endeavors of entity linking focus on linking entities from unstructured texts to those in a knowledge graph (KG)~\cite{clark2016deep, Kolitsas2018ACL, yamada2020luke, klie2020zero, hou2020improving, angell2021clustering}.
Recent work has investigated a cross-domain setting which links entities from heterogeneous texts like blogposts or news to a KG. They only train a model on the labeled source domain and directly apply it to different texts~\cite{gupta2017entity,le2018improving, logeswaran2019zero, zemlyanskiy2021docent}. 
\smodel differs them in three aspects: 1) they have explicit labels, 
2) an AMiner expert, consists of multiple papers, is more complex than an entity from the unstructured text,
and 3) the information gap between AMiner and external sources prevents us from directly using the pre-trained model to external sources. 

\vpara{Contrastive Learning.}
Contrastive learning, which learns the data co-occurrence relationships via instance discrimination task~\cite{wu2018unsupervised}, is a label-efficient representation learning regime. It has shown strength in various domains such as natural language process~\cite{devlin2018bert,yang2019xlnet, brown2020language}, computer vision~\cite{chen2020simple, he2019momentum}, and graph~\cite{qiu2020gcc, hu2020gpt}.
Beyond learning representation, we need a fine-grained metric for matching experts.

The expert discrimination task is related to paper clustering, aiming to partition papers into a set of disjoint clusters corresponding to real experts~\cite{zhang2018name,qiao2019unsupervised}, and author identification, assigning the right authors to the anonymous papers~\cite{chen2020conna, wang2020author}. Different from them, we contrast expert instances.

\vpara{Adversarial Domain Adaptation.}
\hide{
\textcolor{red}{
Unsupervised pre-training can make the downstream models robust to target domain data, where BERT~\cite{devlin2018bert} and XLNET~\cite{chen2012marginalized} can be used as the open-corpus pre-training models, \cite{chen2012marginalized,yang2015unsupervised}  pre-train on the source plus target domain unlabeled data, and~\cite{logeswaran2019zero} further combines the open-corpus pre-training, source plus target pre-training and target only pre-training. They all use pre-training to discover universal features across domains.
} \textcolor{blue}{This paragraph is not clear. }
}
Adversarial learning has been extensively studied for the cross-domain transfer~\cite{ganin2016domain, zhai2020ad} such as word translation~\cite{lample2018word}, 
text classification~\cite{guo2020multi} and relation extraction~\cite{shi2018genre} in different domains. Different from prior end-to-end adversarial learning, we adopt it for fine-tuning, which is an independent process following the pre-training stage.


	\section{Problem Formulation}
\label{sec:problem}
This section defines an expert and formulates the problem of zero-shot expert linking in a contrastive way.


\begin{figure*}[t]
	\centering
	\includegraphics[width=0.75\textwidth]{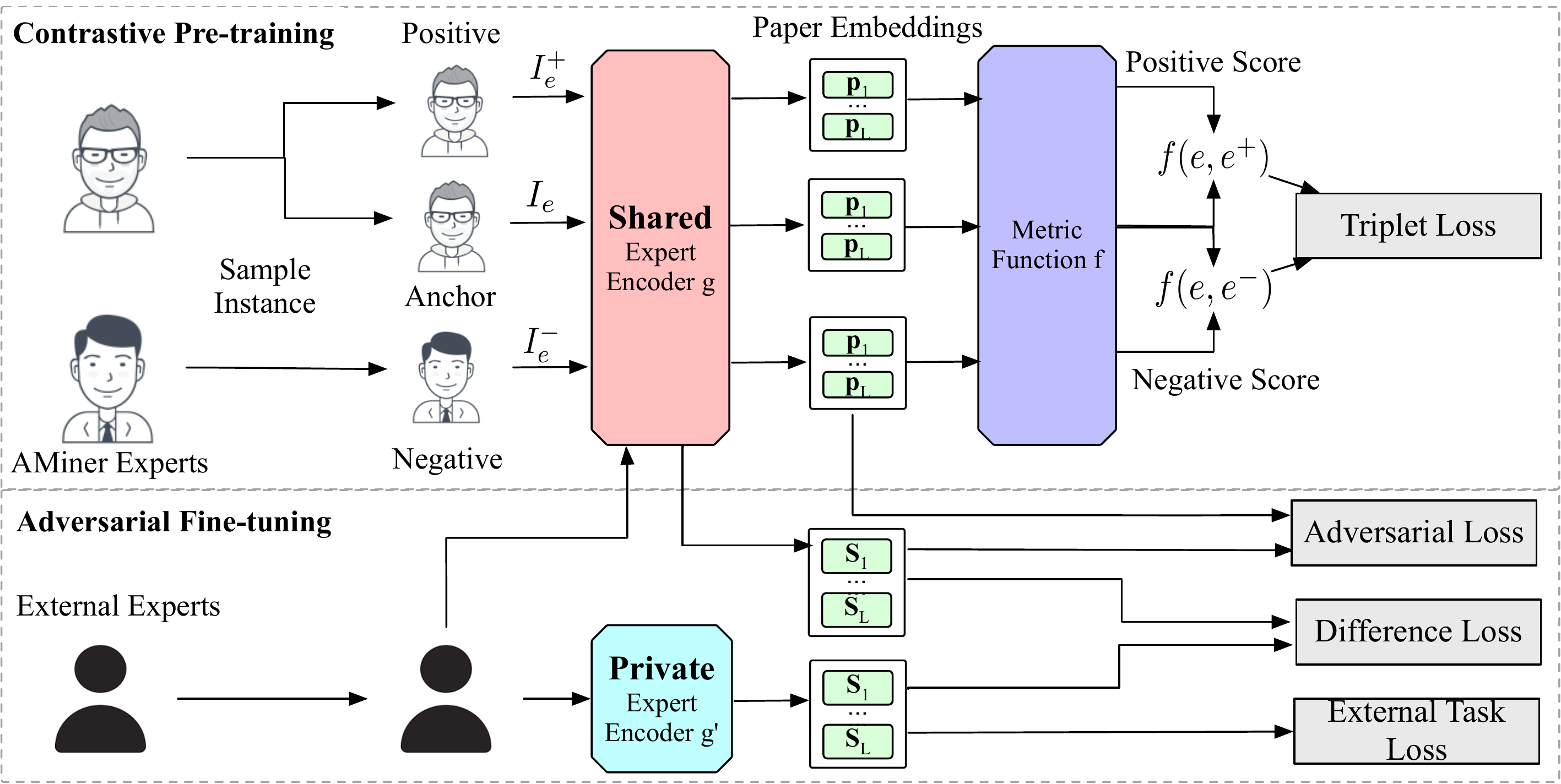}
	\caption{\label{fig:framework} \textbf{Overview.} The pre-training module learns an expert encoder $g$ and a metric $f$ via discriminating the positive expert instance pair $(I_e,I_e^{+})$ from the negative one $(I_e,I_e^{-})$. Then it is fine-tuned on the external experts in an adversarial manner. }
\end{figure*}

\begin{definition}
	\textbf{Expert.} An expert $e$ is comprised by a set of support information $c_e= \{s_1,s_2,\cdots, s_{n_e}\}$, where $s_i$ is a piece of support information. $n_e$ denotes the size of $c_e$.
\end{definition}

The support information varies from different sources. If $e$ is from a news article, $c_e$ can be surrounding texts of the expert name where $s_i$ is one of the sentences.
If $e$ is from LinkedIn, $c_e$ can be a homepage where $s_i$ is one of the attributes such as summary, affiliations, etc. 
Particularly, we denote $c_e$ of an AMiner expert as $c_e = \{p_1, \cdots, p_{n_e}\}$, where $p_i$ is a paper containing title, keywords, venue, etc.




\begin{problem}
	\textbf{Zero-shot Expert Linking.} Given a set of experts $\{e\}$ on AMiner, we aim at pre-training an expert encoder $g$ 
	and a metric function
	$f : \{g(e),g(e')\} \rightarrow \{y\}$ to infer the alignment label between $e$ and $e'$, where $y=1$ implies $e$ and $e'$ are equivalent and $0$ otherwise. After pre-training, we fine-tune $g$ and $f$ on the external experts $\{\tilde{e}\}$ such that $g$ and $f$ are transferred to unseen external sources.
	
\end{problem}

Note that, we assume any experts from external sources can be certainly aligned to the counterparts in AMiner and leave the problem of non-existing alignment to the future.

\hide{
\begin{problem}
\textbf{Domain Adaptation from Aminer to  External Persons.}	
Given an expert encoder $g$ and a metric function $f$ learned on AMiner, a set of AMiner experts $\{e\}$ and external persons $\{a\}$, we are aiming at fine-tuning the expert encoder $g$ and the metric function $f$ to make the representation of any external person $g(a)$ be as close as experts in AMiner, and also make the metric function $g(f(a),f(e))$ be more reliable to capture the matching patterns between $e$ and $a$.	
\end{problem}
}

\section{The \smodel Framework}
\label{sec:approach}



This section introduces \smodel (Figure~\ref{fig:framework}), which consists of a contrastive pre-training module and an adversarial fine-tuning module. The former one pre-trains an expert encoder $g$ and a metric function $f$  purely on AMiner via contrastive learning, and the latter one fine-tunes $g$ and $f$ on both AMiner and the external data in an adversarial manner. For convenience, we hereinafter name the pre-trained model as \pretrain, and also the final fine-tuned model as \model.

\subsection{Contrastive Pre-training Module}
The pre-training module targets at learning an encoder $g$ to capture the universal representation patterns of experts and a metric $f$ to measure the fine-grained matches between experts.
In light of this, we define expert discrimination as our pre-training task to increase the similarities of positive expert instances, while decreasing those of negatives ones. To achieve this, four questions should be answered carefully:


\begin{itemize}
	\item \textbf{Q1:} What is an instance of an expert?
	\item \textbf{Q2:} How to encode an expert instance?
	\item \textbf{Q3:} How to measure the similarity of two instances?
	\item \textbf{Q4:} Which kind of loss function should be selected?
\end{itemize}

\vpara{Q1: An Instance of An Expert.}
we define an instance as a set of randomly sampled papers of the expert $c_e$ on AMiner, i.e. 
an instance $I_{e}$ of $e$ can be formulated as follows:

\beq{
	\label{eq:expert_instance}
	I_{e} = \{p_1, \cdots, p_{L} \},
}

\noindent where $p_i \in c_e$ and $L$ is the maximal number of sampled papers for each instance.  Two instances sampled from the same expert is viewed as a positive pair, while the two instances from different experts are viewed as a negative pair.

\vpara{Q2: BERT-based Expert Encoder.}
Since the support information of experts on multi-sources may be in different languages,  we adopt the multi-lingual BERT~\cite{wolf-etal-2020-transformers} to project the information into a unified semantic space.

In practice, we encode each paper as a basic representation unit of an expert instead of directly encoding the expert as a whole. We concatenate the paper attributes, including title, keywords, venues, etc, as the input $p$ of BERT and apply a Multi-layer Perceptron (MLP) to get the paper embedding:

\hide{
\small
\beq{
	\label{eq:encoder_bak}
	\textbf{p} = \text{MLP}( \text{CLS}(\text{authors}(p))) \oplus \text{MLP}(\text{CLS}(\text{other attributes}(p))),
}
\normalsize
}

\beq{
	\label{eq:encoder}
	\textbf{p} = g(p) = \text{MLP}(\text{CLS}(p)),
}

\noindent where $\text{CLS}(p)$ indicates the CLS token embedding of BERT. 


\hide{
\ipara{\textbf{Representation-based Metric Function.}} 
Representation-based metric function aims to encode each expert into an embedding, and then directly estimate the distance between two experts in the vector space. Specifically, we average the paper embeddings $\{\textbf{p}_n\}_{n=1}^{L}$ of an expert instance $I_e$ to obtain $\textbf{e} $ and measure the Euclidean distance of $\textbf{e}$ and $\textbf{e}'$:

\beq{
	\label{eq:representation-metric-function}
	\textbf{e}  = \frac{1}{L} \sum_{n=1}^{L} \textbf{p}_n, \qquad f(e,e') = \|\textbf{e} - \textbf{e}'\|^2. 
}
}


\hide{
\begin{figure}[t]
	\centering
	\includegraphics[width=0.43\textwidth]{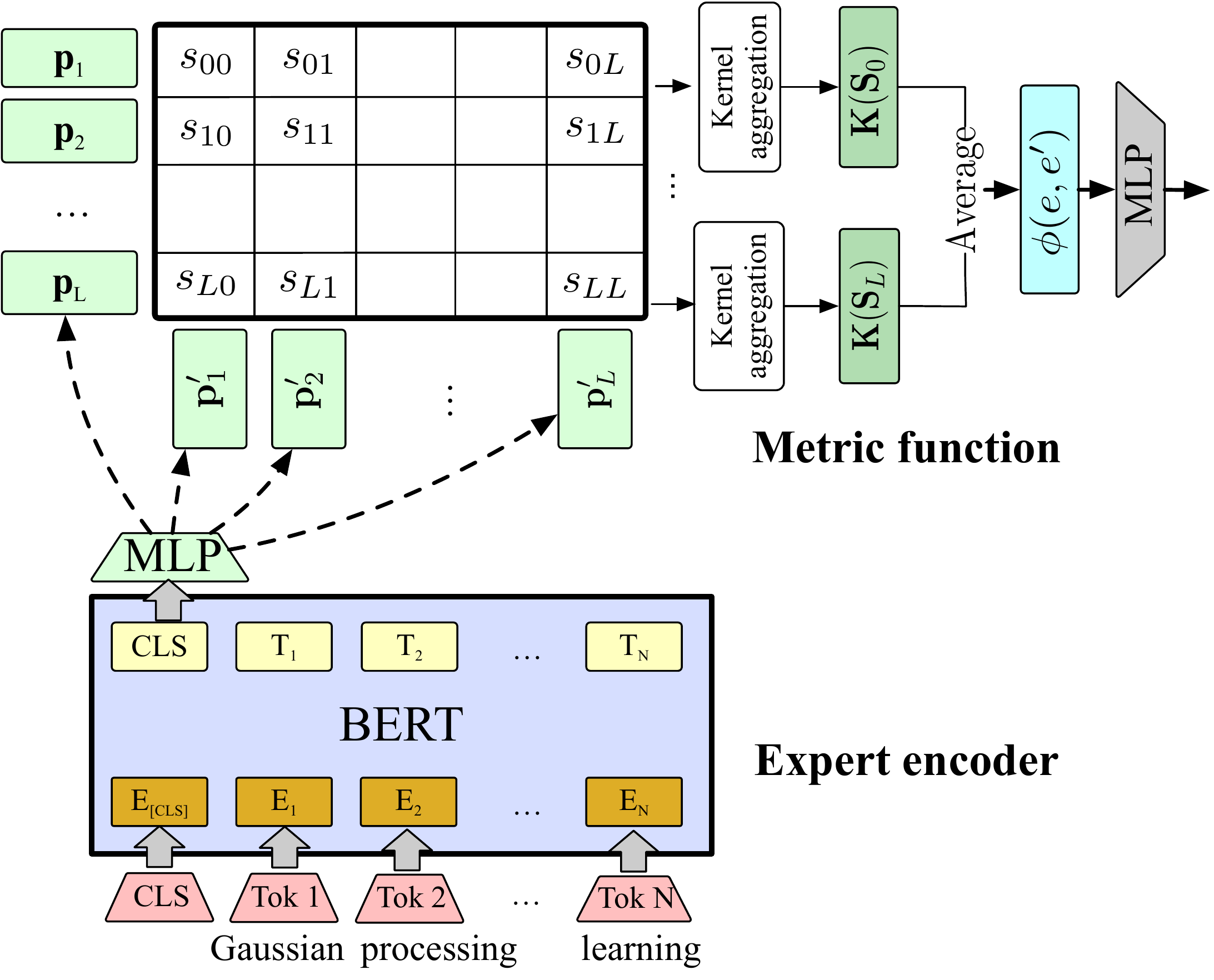}
	\caption{\label{fig:metric_function} The expert encoder and the metric function.}
\end{figure}
}

\vpara{Q3: Interaction-based Metric Function.}
Standard representation-based metric function usually aggregates all the paper embeddings of an instance as the expert embedding, based on which it estimates the similarity of two instances. However, the mixed expert embedding suffers from semantic drift. Thus, inspired by the similar ideas of information retrieval~\cite{xiong2017end, dai2018convolutional}, we propose an interaction-based metric to measure the fine-grained similarity between each paper pair of two instances. 


Formally, we use Eq.\eqref{eq:encoder} to obtain a set of paper embeddings $\{\textbf{p}_m\}_{m=1}^{L}$ and $\{\textbf{p}'_n\}_{n=1}^{L}$ for $I_e$ and $I_{e'}$, respectively. Then we compute a similarity matrix $\mathbf{A}$ between $I_e$ and $I_{e'}$.  Each element $\alpha_{mn}$ is calculated by the normalized euclidean distance  between the $m$-th paper in $I_e$ and $n$-th paper in $I_{e'}$, i.e. $\alpha_{mn} = \|\textbf{p}_m - \textbf{p}'_n\|^2_2$.
Then we adopt an RBF kernel aggregation function~\cite{dai2018convolutional}, which pays attention to how many similar paper pairs within two instances, to extract the similarity patterns. 
Specifically, we transform $\alpha_{mn}$ into a $K$-dimensional distribution (Eq.\eqref{eq:vector}), the $k$-th element of which is converted by the $k$-th RBF kernel with mean $\mu_k$ and  variance $\sigma_k$ (Eq.\eqref{eq:kernel}) implying how likely $\alpha_{mn}$ corresponds to the $k$-th similarity pattern.
Then we sum up over rows to represent the similarities between $m$-th paper in $l_{e}$ and all the papers in $I_{e'}$, and sum up over columns to represent the similarity between $I_e$ and $I_{e'}$ (Eq.\eqref{eq:sum}). Finally, we apply a MLP layer to obtain the similarity score (Eq.\eqref{eq:interaction_metric_function}).

\beqn{
	\label{eq:kernel}
	K_k(\alpha_{mn}) &=&  \exp \left[ -\frac{(\alpha_{mn} - \mu_k)^2}{2\sigma_k^2}\right], \\
	\label{eq:vector}
	\mathbf{K}(\alpha_{mn})&=& \left[ K_1(\alpha_{mn}), \cdots, K_K(\alpha_{mn})\right],  \\
	\label{eq:sum}
	\phi(e, e') &=& \sum_{m=1}^{L} \log \sum_{n=1}^{L}\mathbf{K}(\alpha_{mn}). \\
	\label{eq:interaction_metric_function}
	f(e,e') & =&\text{MLP} (\phi(e, e'))
}

\vpara{Q4: Triplet Loss Function.}
We prefer the triplet loss instead of the widely-used contrastive loss~\cite{chen2020simple} in our problem. Contrastive loss encourages similar instances into a single point in the embedding space~\cite{zhang2018name}. However, experts may publish papers on different topics, making it weird to force all the papers into a single point. Thus the triplet loss, maintaining a relative distance between positive and negative pairs, is a better choice.

Specifically, for the anchor instance $I_{e}$, the positive counterpart $I^{+}_{e}$ is sampled from the same expert, while a negative one $I^{-}_{e}$ is from a different expert. 
Given a set of triplets $\{(I_{e}, I^+_{e},I^-_{e})\}$, the triplet loss function is defined as:

\beq{
	\label{eq:triplet_loss}
	\mathcal L^{\text{pre}}(\theta_g, \theta_f) \!= \!\!\!\!\!\!\!\!\! \sum_{(I_e, I^+_{e},I^-_{e})} \!\!\!\!\!\!\! \max\{0, m + f(e,e^-) - f(e, e^+)\},
}

\noindent where $m$ is a margin, $\theta_{g/f}$ are the parameters of $g$ or $f$. To avoid trivial results, we omit overlaps within instance pairs.

\subsection{Adversarial Fine-tuning Module}
\label{sec:adaption}
Intuitively, \spretrain can be directly applied on external sources for zero-shot expert linking. 
However, the morphology, syntax, topics of the external information may be significantly different from that on AMiner, which encourages us to fine-tune \spretrain to improve its transferability.


Most of the domain adaptation methods assume each domain is comprised of domain-agnostic and domain-private features, thus they learn a shared generator and a private generator for each domain~\cite{liu2017adversarial, shi2018genre}.
However, as our goal is to link external experts to AMiner, we need to extract the features similar to AMiner from external sources as much as possible, such that the pre-trained metric $f$ can better capture the similarity patterns between external and AMiner experts. Inspire by this, besides the shared generators in both domains, we only create a private generator for external experts to get rid of the dissimilar features compared with AMiner, shown in Figure~\ref{fig:framework}. 

\vpara{Generator.}
Besides the shared generator $g^{\text{shared}}$ pre-trained by the pre-training module, we create the same private generator $g^{\text{private}}$  to extract the domain-private features from external experts. 
To enforce the shared and private generator to encode different aspects of features, we adopt orthogonality constraints as a difference loss~\cite{liu2017adversarial}:
\beq{
	\label{eq:orthogonality_constraint}
	\mathcal L^{\text{diff}}(\theta_{g}^{\text{shared}}, \theta_{g}^{\text{private}}) = \sum_{i=1}^{N_{\text{ext}}}||g^{\text{shared}}(s_i)^\text{T} g^{\text{private}}(s_i)||_F^2,
}

\noindent where $||.||_F^2$ is the squared Frobenius norm, 
$N_{\text{ext}}$ is  the number of pieces of the support information $s_i$.

\vpara{Domain Discriminator.}
To cripple the external private features from the shared feature space, we design a domain discriminator for enforcing $g^{\text{shared}}$ to abandon the private features from external sources. 
Given $s_i$ from either AMiner or the external source, we use the shared generator $g^{\text{shared}}$  to extract its features, and apply a classifier $h$ to predict whether it is from the external source or AMiner. We adopt Gradient Reversed Layer~\cite{ganin2016domain} to confuse $h$ such that it cannot distinguish the source of support information:
\beqn{
	\label{eq:adversarial}
	\mathcal L^{\text{adv}}(\theta^{\text{shared}}_g, \theta_h) \!\!\!&=&\!\!\!\!\!\! \sum_{i = 0}^{N_{\text{AMiner}}} \!\!\!\! \log(\hat{p}_i) +\sum_{i = 0}^{N_{\text{ext}}} \log(1-\hat{p_i}), \\ \nonumber
	\hat{p}_i &=& h \left(g^{\text{shared}}(s_i) \right) = \text{MLP}\left(g^{\text{shared}}(s_i) \right),
}

\noindent where $\hat{p}_i$ denotes the likelihood of $s_i$ from AMiner,  $N_{\text{AMiner}}$ is the number of papers on AMiner, h is an MLP layer.
        
\vpara{Task Predictor.}
Likewise, we design an external task predictor $\tilde{h}$, which predicts the source of private features, to further prevent the private features into the share feature space. 


\begin{small}
\beq{
	\label{eq:ext_task_predictor}
	\mathcal{L}^{\text{ext}}(\theta^{\text{private}}_g, \theta_{\tilde{h}}) \!=\! -\sum_{i =0}^{N_{\text{ext}}} \log (1- \hat{p}_i), \quad \hat{p}_i = \tilde{h} (g^{\text{private}}({s}_i))
}
\end{small}

\noindent where $\hat{p}_i$ denotes the probability of the support information $s_i$ is from AMiner. The classifier $\tilde{h}$ is the same as Eq.\eqref{eq:adversarial}.
\subsection{Training and Inference}
The final loss function is defined as follows:

\begin{small}
\beq{
	\label{eq:fine-tune}
	\mathcal L(\theta^{\text{shared}}_g, \theta^{\text{private}}_g, \theta_f, \theta_h) = \mathcal L^{\text{pre-train}} + \alpha \mathcal L^{\text{adv}} + \beta \mathcal L^{\text{diff}} + \gamma \mathcal L^{\text{ext}},
}
\end{small}

\noindent where $\alpha$, $\beta$, $\gamma$ are trade-off hyper-parameters. 
During training, we first pre-train an expert encoder $g^{\text{shared}}$ and a metric$f$ on AMiner via Eq.\eqref{eq:triplet_loss}. Then we fine-tune $g^{\text{shared}}$ and $f$ by adversarial learning via  Eq.\eqref{eq:fine-tune}.
During inference, we use the fined-tuned model to perform zero-shot expert-linking between AMiner and external sources.



\section{Experiments}
\label{sec:setting}

In this section, 
we first evaluate the representation and matching capacity of the pre-trained model \spretrain by two intrinsic tasks, author identification and paper clustering, on AMiner.  Then we validate the transferability of fine-tuned model \smodel by external expert linking, i.e., linking experts from the news article or LinkedIn to AMiner experts. For each experiment, we run 5 trials and report the mean results. Experimental details please refer to Appendix. 

\subsection{Datasets.} 
Table~\ref{tb:dataset} summarizes statistics of three datasets.

\begin{itemize}[leftmargin=*]
	\item \textbf{AMiner.} We employ WhoIsWho\footnote{https://www.aminer.cn/whoiswho}, the largest manually-labeled name disambiguation dataset collected from AMiner, as the basis to be aligned. 

    \item \textbf{News.} We collect news articles from several Chinese technique platforms such as sciencenet, jiqizhixin, etc, 
	and extract names from these news articles by NER tools\footnote{http://thulac.thunlp.org/}, then we link them to AMiner experts by a majority voting of three professional annotators' results. 
		
	\item \textbf{LinkedIn.} We adopt the dataset from~\cite{zhang2015cosnet}. 
	
	
\end{itemize}

\begin{table}
	\newcolumntype{?}{!{\vrule width 1pt}}
	\newcolumntype{C}{>{\centering\arraybackslash}p{3em}}
	\caption{
		\label{tb:dataset} \textbf{Data statistics.} 
		The support information of AMiner, News, and Linkedin are papers, news articles, and  homepages, respectively. \#Avg. candidates are the average number of candidate experts on AMiner for an author in the paper, a name in the news article, or a LinkedIn user.
	}
	\centering 
	\footnotesize
	\renewcommand\arraystretch{1.0}
	\begin{tabular}{@{~}l@{~}?*{1}{CCC}@{~}}
		\toprule
		& AMiner & News & LinkedIn \\
		\midrule
		\#Experts	                  &  45,187   &  1,824  &  1,665         \\
		\#Support Information 	&  399,255 &  20,658  & 50,000 \\
		\#Avg. candidates          &  18         & 8.79    &4.85      \\
		
		\bottomrule
	\end{tabular}
	
\end{table}

\vpara{Candidates.}
Given an external expert to be linked, we choose the AMiner experts with similar names as candidates. The similar names are obtained by moving the last name to the first or keeping the name initials except for the last name. For example, the similar names of ``Bo Li" include ``Li Bo", ``B Li" and ``L Bo". 

\hide{
we follow the task of author identification~\cite{chen2017task,zhang2018camel}, which targets at linking the right authors for a given paper, to evaluate the pre-training module on a sampled dataset of AMiner.
More training and test setting details can be found in the Appendix.
We estimate the similarity score between a target paper and each candidate expert by the metric function of each method and rank all the candidate experts by the similarity scores.  We use the Hit Ratio before top K ( HR@K, K=1,3) to measure the proportion of the correctly identified experts ranked in top K candidates, and use MRR to measure the average reciprocal ranks of the correctly identified experts. 
}

\subsection{Evaluation of the Pre-training Module} 
We adopt two intrinsic tasks, author identification~\cite{chen2020conna, wang2020author} and paper clustering~\cite{zhang2018name}, on AMiner, to evaluate the representation and matching capacity of \spretrain via contrastive pre-training. 

\subsubsection{Author Identification} assigns a new paper to the right expect, following the second track of competition\footnote{https://biendata.xyz/competition/aminer2019\_2/}. Thus we adopt the same test set.
We estimate the similarity score between a new paper and each candidate via (Eq.\eqref{eq:interaction_metric_function}),
and return the expert with the highest similarity as the right answer.

 \vpara{Evaluation Metrics.} We use HitRatio@K (HR@K, K=1,3) to measure the proportion of correctly assigned experts ranked in top K, and use MRR to measure the average reciprocal ranks of correctly assigned experts. 

\vpara{Baselines.} 
We employ the following methods to solve the task of author identification:  \textbf{GBDT}~\cite{li2013feature} is a  feature-engineering model in KDD Cup 2013~\cite{roy2013microsoft},	\textbf{Camel}~\cite{zhang2018camel} represents a paper by GRU with its title and keywords, an expert by one-hot embedding. 
\textbf{HetNetE}~\cite{chen2017task} is similar to Camel except that each paper is represented by the author names, affiliations, venues in addition to the title and keywords,
and \textbf{CONNA}~\cite{chen2020conna} is an interaction-based model. The basic interactions are built between the token embeddings of two attributes, then different attributes matrices are aggregated as the paper-level interactions, finally the paper-level matrices are aggregated as expert-level interactions.

\begin{table}
	\newcolumntype{?}{!{\vrule width 1pt}}
	\newcolumntype{C}{>{\centering\arraybackslash}p{3em}}
	\caption{
		\label{tb:author_identification} Performance of Author Identification on AMiner.
	}
	\footnotesize
	\centering 
	\renewcommand\arraystretch{1.0}
	\begin{tabular}{l?ccc}
		\toprule
		Model & {HR@1} & {HR@3} & {MRR} \\
		\midrule

		GBDT
		& 0.873 & 0.981 & 0.927 \\	
		
		Camel 
		& 0.577  & 0.737&  0.644 \\
		
		HetNetE 
		& 0.582 & 0.759 & 0.697 \\

		CONNA
		& \textbf{0.911} & \textbf{0.985} & \textbf{0.949}\\
		
		\midrule
		Unsupervised
		& 0.713   & 0.875    & 0.808     \\
		Paper-paper pseudo labels
		& 0.864  & 0.960    &  0.915 \\
		Paper-expert pseudo lables
		& 0.892  & 0.970    & 0.933  \\
		Representation metric function 
		& 0.870    &  0.956     & 0.918   \\
		\midrule
		\pretrain  
		& 0.898    &  0.964     &   0.934   \\
		
		\bottomrule
	\end{tabular}
	
\end{table}

\vpara{Results.}
Table~\ref{tb:author_identification} shows the performance of author identification. 
We can see \spretrain outperforms Camel and HetNetE by 31.6$\sim$32.1\% in HR@1. 
Camel and HetNetE adopt a representation-based metric function, which embed a  paper and an expert independently, and computes the similarity score between them. 
Thus they fail to capture the fine-grained similarities between papers. 

\spretrain is comparable with GBDT and CONNA. GBDT, extracts hand-crafted features between a paper and an expert, and CONNA, computes the fine-grained interactions between them, outperform Camel and HetNetE by a large margin, which demonstrates the efficacy of interaction-based strategy. CONNA slightly outperforms \spretrain by 1.3\% in HR@1, because CONNA, with the dedicated architecture and training criteria, is tailed for the author identification task. 
However, GBDT and CONNA
aren't capable of being transferable to other domains, while \spretrain first applies the expert encoder $g$ to encode each paper into a basic embedding unit, and uses a metric $f$ to compute the similarities between papers and experts. The $g$ and $f$ can be easily generalized to other domains, such as news articles or LinkedIn, by first encoding support information with $g$ and then estimate the interactions with $f$ similarly. 

\begin{table}
	\newcolumntype{?}{!{\vrule width 1pt}}
	\newcolumntype{C}{>{\centering\arraybackslash}p{3em}}
	\caption{
		\label{tb:paper-clustering} Performance of Paper Clustering on AMiner.
	}
	\footnotesize
	\centering 
	\renewcommand\arraystretch{1.0}
	\begin{tabular}{l?ccc}
		\toprule
		Model & {P-Pre.} & {P-Rec.} & {P-F1} \\
		\midrule
		
		Louppe et al.
		& 0.609  & 0.605 &  0.607 \\
		Zhang et al. 
		& 0.768 & 0.551 & 0.642 \\
		G/L-Emb 
		& \textbf{0.835} & 0.640 & 0.724 \\
		
		\midrule
		Unsupervised
		& 0.332   & 0.591    & 0.425     \\
		Paper-paper pseudo labels
		& 0.659  & 0.779    &  0.714 \\
		Paper-expert pseudo lables
		&  0.715   & 0.786    & 0.749  \\
		Representation metric function 
		& 0.595    &  0.754     & 0.665   \\
		\midrule
		\pretrain  
		& 0.724   &  \textbf{0.789}     &  \textbf{0.755}  \\
		
		\bottomrule
	\end{tabular}
	
\end{table}

\subsubsection{Paper Clustering} aims at clustering papers belonging to the same expert together, following the first track of the competition\footnote{https://biendata.xyz/competition/aminer2019/}. 
We adopt the same test set, and use the hierarchical agglomerative clustering algorithm (HAC) to cluster papers based on the paper embeddings output by $g$. 

\vpara{Evaluation Metrics.} We use pairwise Precision, Recall, and F1-score (P-Pre., P-Rec, and P-F1)~\cite{zhang2018name}  to evaluate the clustering results of each name,  and then calculate the macro metric by averaging metrics over all names.

\vpara{Baselines.} We compare with three state-of-the-art methods for paper clustering. 
\textbf{Louppe et al.}~\cite{louppe2016ethnicity} trains a similarity metric based on hand-crafted features to measure the similarities between papers, 
\textbf{Zhang et al.}~\cite{zhang2017name} constructs three graphs including the expert-expert graph, the expert-paper graph, and the paper-paper graph for each name. Then they learn graph embedding by preserving node connectivity on all the graphs,
and
\textbf{G/L-Emb}~\cite{zhang2018name} learns paper embeddings on a global paper-paper network, and then fine-tunes the paper embeddings on a local paper-paper network built for each name by graph auto-encoding. 

\hide{
\begin{itemize}[leftmargin=*]
	\item 
	\textit{\textbf{Louppe et al.}}~\cite{louppe2016ethnicity}: trains a similarity function based on hand-crafted features to directly measure the similarities between papers.
	
	\item 
	\textit{\textbf{Zhang et al.}}~\cite{zhang2017name}: constructs an expert-expert graph (coauthors are linked), an expert-paper graph (a paper is linked to its author) and a paper-paper network (papers coauthored by the same author are linked) for each name. Then they learn graph embedding by preserving node connectivity on each network. 
	
	\item
	\textit{\textbf{G/L-Emb}}~\cite{zhang2018name}: first learns paper embeddings on a global paper-paper network 
	and then fine-tunes the paper embeddings on a local paper-paper network built for each name by graph auto-encoding. 

\end{itemize}
}
\vpara{Results.}
Table~\ref{tb:paper-clustering} presents the performance of paper clustering. 
Overall, \spretrain advances other baselines by 3.1$\sim$14.8\% in Pairwise-F1. Among them, Louppe et al. perform the worst, as it merely captures the pairwise similarities while ignoring the interplays with other papers.
Zhang et al. and G/L-Emb, which build the local paper-paper graph in each name and leverage the graph structure to learn paper embeddings, outperform Louppe et al. by 3.5$\sim$11.7\% in Pairwise-F1.
Besides local embeddings, G/L-Emb incorporates global information via preserving the connectivity in the global graph, making it outperform Zhang et al by +8.2\% in Pairwise-F1. 
Although \spretrain discards the graph information for its limitation in transferability, it still performs the best because of the striking representation capability endowed by the BERT encoder $g$ and contrastive pre-training.


\begin{figure}
	\centering
	
	\subfigure[Paper Sampling.]{\label{subfig:instances}
		\includegraphics[width=0.22\textwidth]{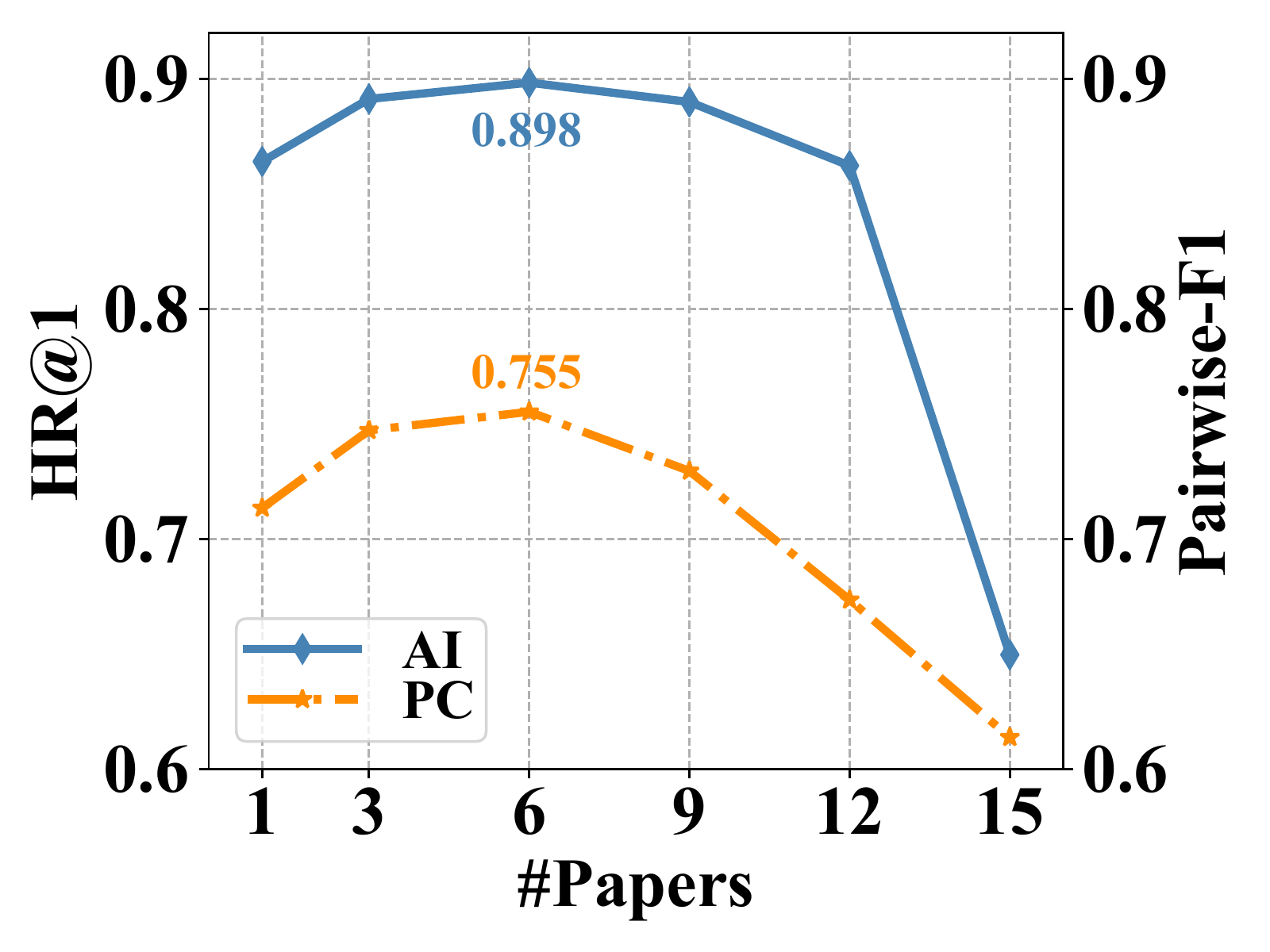}
	}
	\hspace{-0.05in}
	\subfigure[Negative Sampling.]{\label{subfig:negatives}
		\includegraphics[width=0.22\textwidth]{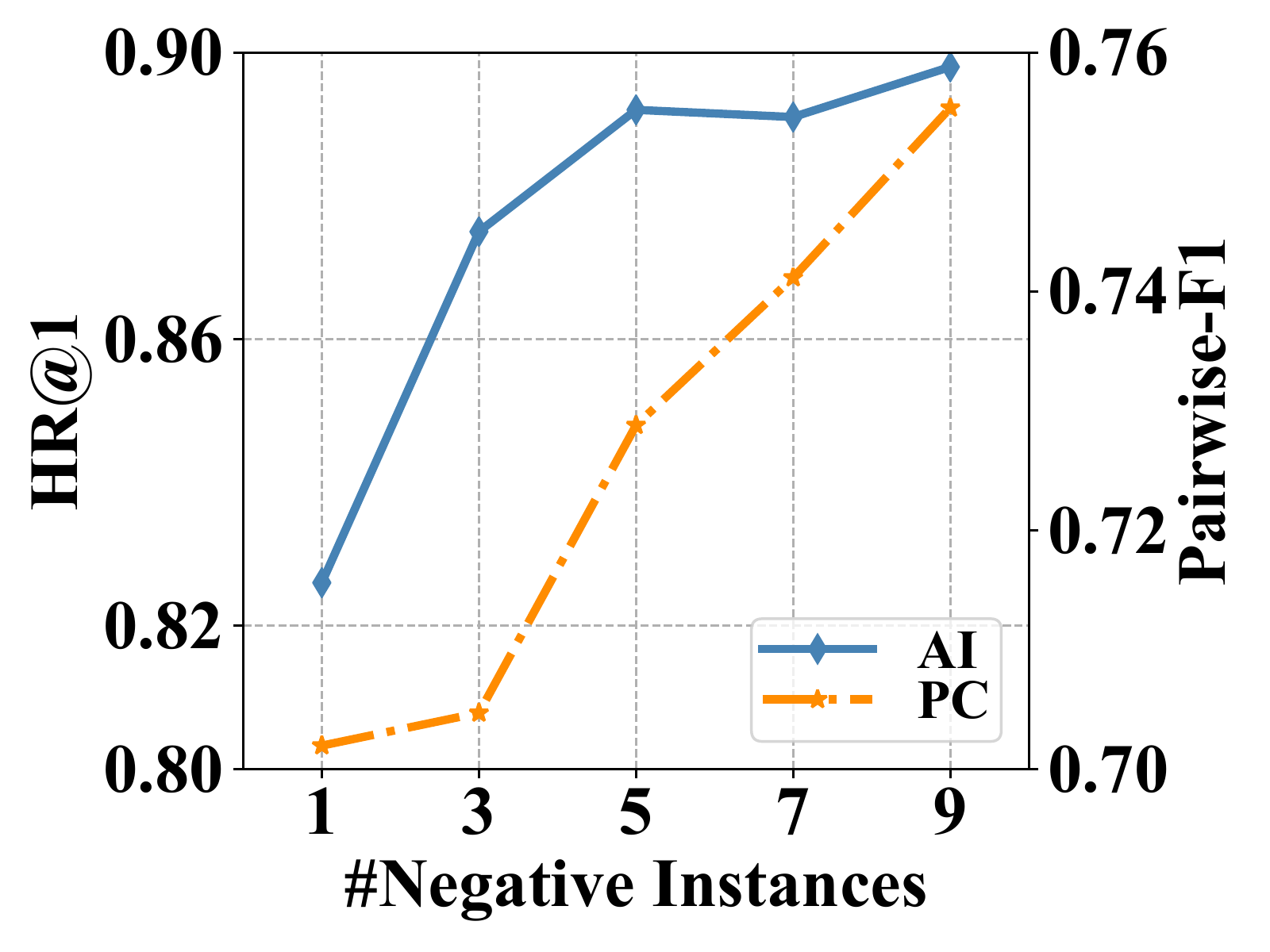}
	}

	\caption{\label{fig:para} The Effect of (a) paper sampling and (b) negative sampling. AI and PC stand for author identification and paper clustering respectively.}
\end{figure}

\subsubsection{Ablation Study} 
\label{sec:insight}

To verify the efficacy of different components in \pretrain, we make four model variants: \textbf{Unsupervised} mimics the unsupervised industrial methods. We use BERT to obtain paper embeddings and average them of an expert as the expert embedding. Then we calculate the euclidean distance between the paper and expert,
\textbf{Paper-paper pseudo labels} samples paper-paper pseudo labels instead of the pairs of instances, likewise two papers are viewed as a positive pair if they are sampled from the same expert and a negative one otherwise,
\textbf{Paper-expert pseudo labels} samples paper-expert pseudo labels,
and \textbf{Representation metric function} uses the representation-based metric function to replace Eq.~\eqref{eq:interaction_metric_function} via averaging all the paper embeddings of an instance as the expert embedding.

\hide{
\begin{itemize}[leftmargin=*]
	\item 
	\textit{Unsupervised:} mimics the unsupervised industrial methods. We use BERT to obtain paper embeddings and average them of an expert as the expert embedding. Then we calculate the euclidean distance between them. 
	
	\item 
	\textit{Paper-paper pseudo labels:} samples paper-paper pseudo labels instead of the pairs of expert instances, likewise two papers are viewed as a positive pair if they are published by a same expert and a negative one otherwise.

	\item
	\textit{Paper-expert pseudo labels:} samples paper-expert pseudo labels instead of pairs of expert instances.
	
	\item
	\textit{Representation metric function:} uses the representation-based metric function to replace Eq.~\eqref{eq:interaction_metric_function} via averaging paper embeddings of an instance as the expert embedding.

	
\end{itemize}
}
The performance is shown in Table~\ref{tb:author_identification} and Table~\ref{tb:paper-clustering}. The unsupervised model performs the worst, -18.5\% in HR@1 and -33.0\% in Pairwise-F1 compared with \pretrain, which implies that the vanilla BERT fails to measure semantic correlations between papers or experts. 
Both the paper-paper and paper-expert pseudo labels underperform \pretrain, -0.6$\sim$3.4\% in HR@1 and -0.6$\sim$4.1\% in Pairwise-F1, denotes contrasting two expert instances can result in better representations of papers and experts. 
The representation-based metric function also underperforms \pretrain, -2.8\% in HR@1 and -9.0\% in Pairwise-F1, which emphasizes the superiority of interaction-based matching strategy.

\vpara{Effect of Paper Sampling.}
We explore how the maximal number $L$ (\#Papers) of an instance affects performance. 
We vary $L$ from 1 to 15 with interval 3 and present the performance of both tasks in Figure \ref{subfig:instances}. We see that 
either too few or many papers will harm the performance. 
Few papers may result in dissimilar expert instances even if they are sampled from a same expert,
while too many papers make positive expert instances easier to be distinguished, thus lead to a trivial solution and degrade the performance~\cite{you2020graph}.



\vpara{Effect of Negative Sampling.}
The number of negative instances also influences the performance. Many efforts~\cite{he2019momentum, chen2020simple} have shown that more negative instances result in better performance. To verify this, we vary negative instances from 1 to 9 with interval 2 in both tasks. From Figure \ref{subfig:negatives}. We can see \spretrain improves incessantly with the increment of negative instances. 
Due to the limitation of GPUs, the maximal number of negative instances is 9. But it is possible to improve the performance when GPUs with larger memory size are available.

\subsection{Evaluation of the Adversarial Fine-tuning Module} 
\label{sec:adversarial}
We fine-tune \spretrain between AMiner and external sources, and evaluate \smodel on external expert linking task.

\vpara{Baselines.}
We compared \smodel with five classical domain adaptation baselines as: \textbf{Unsupervised} is the same as the unsupervised model variant for evaluating \pretrain,
\textbf{\pretrain} directly encodes and links the external experts to AMiner,
\textbf{Chain Pre-training}~\cite{logeswaran2019zero} chains a series of pre-training stages together. We first pre-train the vanilla BERT on both AMiner and the external data, then fine-tune it on the external data to obtain the fine-tuned BERT, base on which we train \spretrain on AMiner via Eq.~\eqref{eq:triplet_loss},
\textbf{DANN}~\cite{ganin2016domain} extracts domain-agnostic without domain-private features. We only use a shared generator to encode both AMiner and external expert,
and \textbf{ASP-MTL}~\cite{liu2017adversarial} captures both the domain-shared and domain-private features. We create private generators for both AMiner and the external source.

\hide{
\begin{itemize}[leftmargin=*]
	\item 
	\textit{\textbf{Unsupervised:}} is the same as the unsupervised variant model for evaluating the pre-training module. 
	
	\item 
	\textit{\textbf{\spretrain}}: directly applies it to encode and link the external persons to AMiner experts.
	\item 
	\textit{\textbf{Chain Pre-training}}~\cite{logeswaran2019zero}: chains together a series of pre-training stages for domain adaptation. We first pre-train the vanilla BERT on both AMiner and the external data, then fine-tune it only on the external data to obtain the fine-tuned BERT, base on which we train \spretrain on AMiner.
	
	\item 
	\textit{\textbf{DANN}}~\cite{ganin2016domain}: extracts domain-agnostic without domain-private features. We only use a shared generator to encode both AMiner and external experts.
	
	\item
	\textit{\textbf{ASP-MTL}}~\cite{liu2017adversarial}: captures both the domain-shared and domain-private features. We create private generate for both AMiner and the external source.
	
	
	
\end{itemize}
}

\begin{table}
	\newcolumntype{?}{!{\vrule width 1pt}}
	\newcolumntype{C}{>{\centering\arraybackslash}p{2em}}
	\caption{
		\label{tb:domain_adaptation} Performance of External Expert Linking.
	}
	\centering 
	\footnotesize
	\renewcommand\arraystretch{1.0}
	\begin{tabular}{@{~}l?@{~}*{1}{CCC?}*{1}{CCC}@{~}}
		\toprule
		
		\multirow{2}{*}{\vspace{-0.3cm} External Sources}
		&\multicolumn{3}{c?}{News}
		&\multicolumn{3}{c}{LinkedIn} 
		
		\\
		\cmidrule{2-4} \cmidrule{5-7} 
		& {HR@1} & {HR@3} & {MRR} & {HR@1} & {HR@3} & {MRR}   \\
		\midrule		
		Unsupervised
		& 0.329 & 0.731 & 0.559  & 0.805 & 0.963& 0.886  \\
		\spretrain
		& 0.737 & 0.927 & 0.837  & 0.897  & 0.982 &  0.940     \\
		Chain Pre-training
		&   0.739     &    0.927       &    0.839      &  0.895 &  0.978&  0.939  \\
		DANN 
		&  0.743   &0.928     &    0.842     & 0.901 & 0.983 &  0.943  \\
		ASP-MTL
		&  0.746 & 0.930      &     0.843     & 0.903 & 0.983 &  0.944 \\
		
		\midrule
		\model
		
		&\textbf{0.753}    & \textbf{0.936} &  \textbf{0.848} & \textbf{0.904} &  \textbf{0.987} &  \textbf{0.945}  \\
		\midrule
		\smodel w/o $\mathcal L^{\text{adv}}$
		&0.721    & 0.923 & 0.825 & 0.873 &  0.981 &  0.927 \\
		\smodel w/o $\mathcal L^{\text{diff}}$
		&0.745    & 0.927 & 0.841 & 0.901 &  0.985 &  0.942 \\
		\smodel w/o $\mathcal L^{\text{ext}}$
		&0.743    & 0.925 & 0.84 & 0.898 &  0.983 &  0.941 \\	
		\bottomrule
	\end{tabular}
	
\end{table}

\vpara{Results.}
Table~\ref{tb:domain_adaptation} shows the performance of linking external experts from news articles or LinkedIn users to AMiner experts. 
The unsupervised model performs the worst (-9.9$\sim$42.4\% in HR@1), as the vanilla BERT isn't fine-tuned at all. 
\pretrain, pre-trained on AMiner without fine-tuning on the external data, underperforms \smodel by 0.7$\sim$1.6\%.
The Chain Pre-training only fine-tunes BERT encoder $g$ but not metric $f$.
Although DANN fine-tunes both $g$ and $f$, it does not detach the private features from the shared space. ASP-MTL, builds both the private and shared encoders, outperforms the above two baselines. 
Compared with ASP-MTL, \smodel performs better (+0.1$\sim$0.7\% in HR@1). As the goal of external expert linking is to extract the features similar to AMiner experts from external sources as much as possible, there is no need to extract the private features of AMiner. Moreover, we create several constraints to make the external private features special enough to be identified.

\hide{
\begin{figure}[t]
	\centering
	\includegraphics[width=	0.47\textwidth]{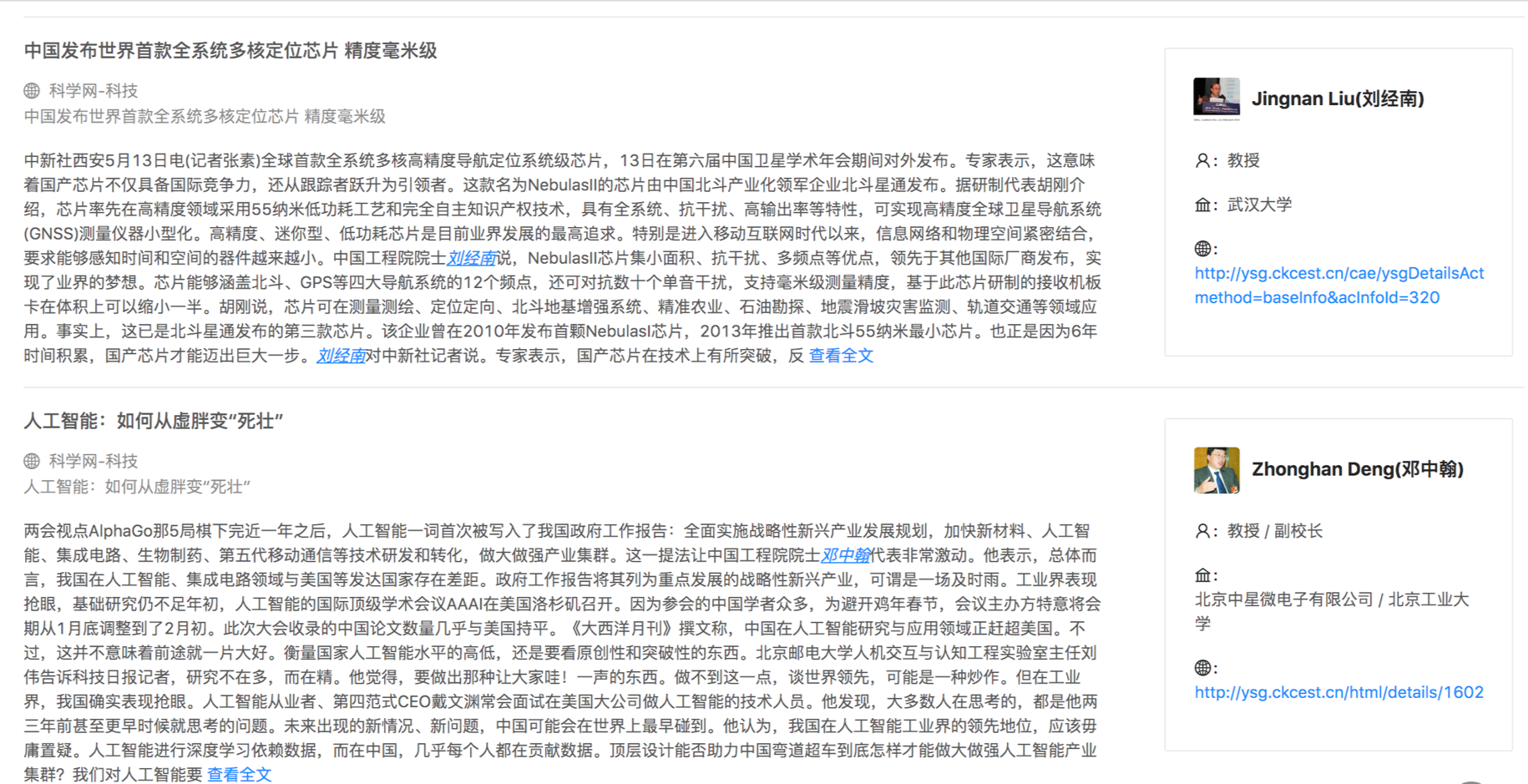}
	\caption{\label{fig:demo} Demo of linking news to AMiner experts.}
\end{figure}
}

\vpara{Ablation Study.} 
To verify the efficacy of different components in \model, we make three model variants: 	\textbf{\smodel w/o $\mathcal L^{\text{diff}}$} removes the difference loss in Eq.\eqref{eq:orthogonality_constraint}, 
\textbf{\smodel w/o $\mathcal L^{\text{adv}}$} removes the adversarial loss in Eq.\eqref{eq:adversarial}, 
and \textbf{\smodel w/o $\mathcal L^{\text{ext}}$:} removes the external task predictor loss in Eq.\eqref{eq:ext_task_predictor}.

\hide{
    \begin{itemize}[leftmargin=*]
    	\item 
    	\textit{\smodel w/o $\mathcal L^{\text{diff}}$:} removes the difference loss in Eq.\eqref{eq:orthogonality_constraint}.
    	\item 
    	\textit{\smodel w/o $\mathcal L^{\text{adv}}$:} removes the adversarial loss in Eq.\eqref{eq:adversarial}.
    	
    	\item 
    	\textit{\smodel w/o $\mathcal L^{\text{ext}}$:} removes the predictor loss in Eq.\eqref{eq:ext_task_predictor}.
    	
    	
    \end{itemize}
}

The performance is shown in Table~\ref{tb:domain_adaptation}.
\smodel w/o $\mathcal L^{\text{adv}}$ performs the worst (-3.1$\sim$3.2\% in HR@1 compared with \model) denoting the domain discriminator takes the most important role in the fine-tuning process. Both \smodel w/o $\mathcal L^{\text{diff}}$ and \smodel w/o $\mathcal L^{\text{ext}}$ underperform \smodel by 0.3$\sim$1.0\% in HR@1, which shows that both losses can enhance model performance. We also observe a desired phenomenon that \smodel w/o $\mathcal L^{\text{diff}}$ performs slightly better than \smodel w/o $\mathcal L^{\text{ext}}$, 
as we expect more about explicitly preventing external private features into the shared feature space.

\begin{figure}
	\centering
	
	\subfigure[Before fine-tuning.]{\label{subfig:before_fine_tune}
		\includegraphics[width=0.21\textwidth]{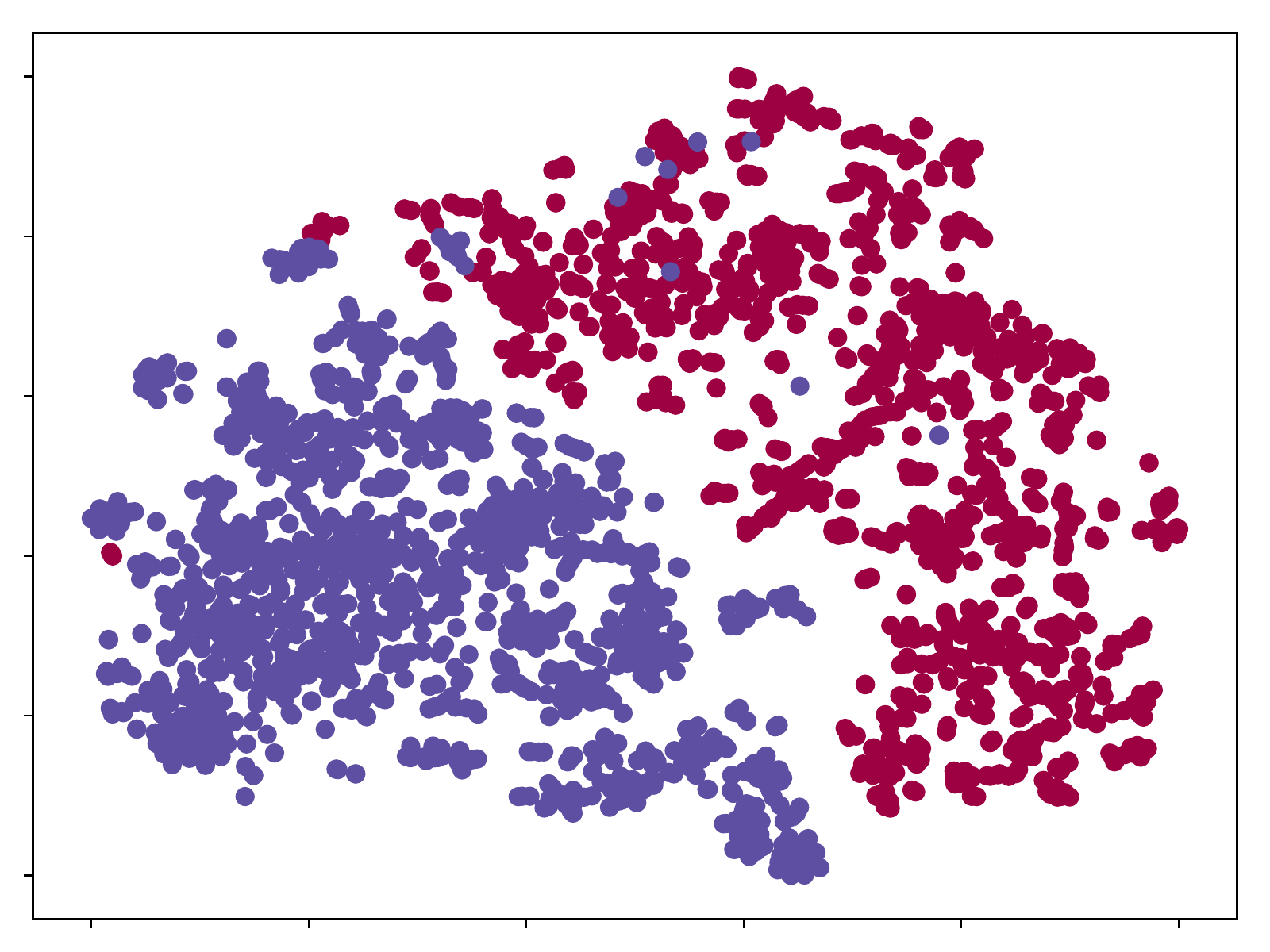}
	}
	\hspace{-0.05in}
	\subfigure[After fine-tuning.]{\label{subfig:after_fine_tune}
		\includegraphics[width=0.21\textwidth]{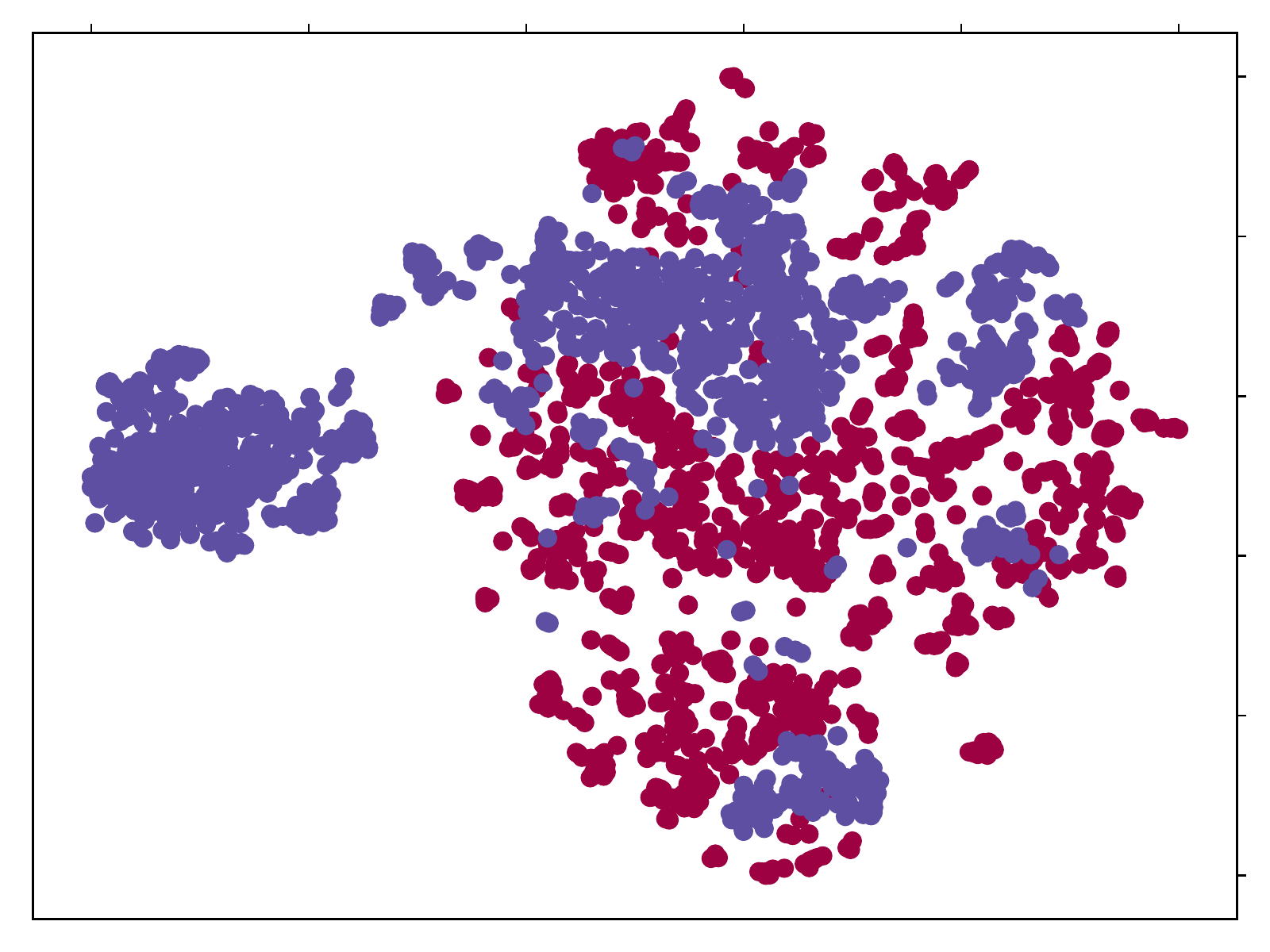}
	}
	
	\caption{\label{fig:visual} T-SNE visualization before and after the adversarial fine-tuning for news articles (\textit{Purple}) and AMiner (\textit{Red}).}
\end{figure}

\vpara{Visualization.}
We utilize t-SNE~\cite{maaten2008visualizing} to visualize papers on AMiner and sentences on news articles output by the encoder $g$ before and after fine-tuning. \figref{fig:visual}(a) implies when using \pretrain, the embedding distributions between AMiner and news articles are not aligned, whille \figref{fig:visual}(b) shows \model, after fine-tuning, mitigates the domain shift.

\subsection{Online Deployment}
\label{sec:onlineD}
A screenshot of deployed system with \smodel is shown in~\figref{fig:motivation}.  
Practically, we first extract names by NER tools from a news article. Then for each name, 
instead of the experimental candidate selection strategy, we adopt ElasticSearch\footnote{https://www.elastic.co} to perform online fuzzy search. Finally, we apply \smodel to estimate the similarity between each candidate and expert from a news article. To solve the case that an external expert may not be linked to any AMiner experts, we pre-defined a threshold and return the candidate with the highest score exceeding the threshold as right expert on AMiner. 


\vpara{Active Learning.}
Admittedly, assignment errors are inevitable, thus we allow users to provide feedback to our assignment results.
Specifically, if users agree with the results, they can click "Thumbs Up" on the top left of each article shown on~\figref{fig:motivation}, otherwise, they can fill in a feedback form to submit the right experts they think on AMiner. The feedback will be regarded as new training instances to further improve the online performance of \model.

	\section{Conclusion and Discussion}
\label{sec:con} 
\vpara{Ethical Statement.} CODE may violate personal privacy via automatically linking the news from experts' personal life to their professional information on AMiner. To avoid this, we have adopted the following two strategies: (1) we only integrate the external public information instead of the private information from the scientific media such as news.sciencenet.cn and jiqizhixin.com, and from professional social platforms such as LinkedIn; (2) before linking the information online, we send the email to the corresponding experts in AMiner to seek their authorization.

We propose \model, consisting of a contrastive pre-training module and an adversarial fine-tuning module, to link experts from external sources to AMiner in the zero-shot setting. The former one performs the pre-training task of expert discrimination to learn an expert encoder for capturing the universal representation patterns of experts, and an interaction-based metric to characterize fine-grained matches between experts on AMiner. The later one adapts the pre-trained model to the unseen external sources when linking experts from them to AMiner in an adversarial manner. 
Experimental results connote the superiority of \model.
In the future, we plan to generalize \smodel into more external sources and deploy it online. 

	\section{Acknowledgments}
We would like to thank Lingxi Zhang and Haipeng Ding for their help in data collection. We also thank the involved staffs in Zhipu.AI team for their help in online deployment of CODE. 
This work is supported by National Natural Science Foundation of China 62076245 and 61825602.
	
	\bibliography{reference} 

\begin{thebibliography}{42}
\providecommand{\natexlab}[1]{#1}

\bibitem[{Angell et~al.(2021)Angell, Monath, Mohan, Yadav, and
  McCallum}]{angell2021clustering}
Angell, R.; Monath, N.; Mohan, S.; Yadav, N.; and McCallum, A. 2021.
\newblock Clustering-based Inference for Biomedical Entity Linking.
\newblock In \emph{Proceedings of the 2021 Conference of the North American
  Chapter of the Association for Computational Linguistics: Human Language
  Technologies}, 2598--2608.

\bibitem[{Brown et~al.(2020)Brown, Mann, Ryder, Subbiah, Kaplan, Dhariwal,
  Neelakantan, Shyam, Sastry, Askell et~al.}]{brown2020language}
Brown, T.~B.; Mann, B.; Ryder, N.; Subbiah, M.; Kaplan, J.; Dhariwal, P.;
  Neelakantan, A.; Shyam, P.; Sastry, G.; Askell, A.; et~al. 2020.
\newblock Language models are few-shot learners.
\newblock \emph{arXiv preprint arXiv:2005.14165}.

\bibitem[{Chen et~al.(2020{\natexlab{a}})Chen, Zhang, Tang, Cai, Wang, Zhao,
  Chen, and Li}]{chen2020conna}
Chen, B.; Zhang, J.; Tang, J.; Cai, L.; Wang, Z.; Zhao, S.; Chen, H.; and Li,
  C. 2020{\natexlab{a}}.
\newblock CONNA: Addressing Name Disambiguation on The Fly.
\newblock \emph{IEEE Transactions on Knowledge and Data Engineering}.

\bibitem[{Chen et~al.(2020{\natexlab{b}})Chen, Kornblith, Norouzi, and
  Hinton}]{chen2020simple}
Chen, T.; Kornblith, S.; Norouzi, M.; and Hinton, G. 2020{\natexlab{b}}.
\newblock A simple framework for contrastive learning of visual
  representations.
\newblock \emph{arXiv preprint arXiv:2002.05709}.

\bibitem[{Chen and Sun(2017)}]{chen2017task}
Chen, T.; and Sun, Y. 2017.
\newblock Task-guided and path-augmented heterogeneous network embedding for
  author identification.
\newblock In \emph{Proceedings of the Tenth ACM International Conference on Web
  Search and Data Mining}, 295--304.

\bibitem[{Clark and Manning(2016)}]{clark2016deep}
Clark, K.; and Manning, C.~D. 2016.
\newblock Deep reinforcement learning for mention-ranking coreference models.
\newblock In \emph{Proceedings of the 54th Annual Meeting of the Association
  for Computational Linguistics: Human Language Technologies}.

\bibitem[{Dai et~al.(2018)Dai, Xiong, Callan, and Liu}]{dai2018convolutional}
Dai, Z.; Xiong, C.; Callan, J.; and Liu, Z. 2018.
\newblock Convolutional neural networks for soft-matching n-grams in ad-hoc
  search.
\newblock In \emph{Proceedings of the eleventh ACM international conference on
  web search and data mining}, 126--134.

\bibitem[{Devlin et~al.(2019)Devlin, Chang, Lee, and
  Toutanova}]{devlin2018bert}
Devlin, J.; Chang, M.-W.; Lee, K.; and Toutanova, K. 2019.
\newblock BERT: Pre-training of Deep Bidirectional Transformers for Language
  Understanding.
\newblock In \emph{Proceedings of the 2019 Conference of the North American
  Chapter of the Association for Computational Linguistics: Human Language
  Technologies, Volume 1 (Long and Short Papers)}, 4171--4186.

\bibitem[{Efimov, Silva, and Solecki(2013)}]{efimov2013kdd}
Efimov, D.; Silva, L.; and Solecki, B. 2013.
\newblock KDD Cup 2013-Author-paper identification challenge: Second place
  team.
\newblock In \emph{KDD Cup 2013 Workshop}, 3.

\bibitem[{Ganin et~al.(2016)Ganin, Ustinova, Ajakan, Germain, Larochelle,
  Laviolette, Marchand, and Lempitsky}]{ganin2016domain}
Ganin, Y.; Ustinova, E.; Ajakan, H.; Germain, P.; Larochelle, H.; Laviolette,
  F.; Marchand, M.; and Lempitsky, V. 2016.
\newblock Domain-adversarial training of neural networks.
\newblock \emph{The Journal of Machine Learning Research}, 2096--2030.

\bibitem[{Guo, Pasunuru, and Bansal(2020)}]{guo2020multi}
Guo, H.; Pasunuru, R.; and Bansal, M. 2020.
\newblock Multi-source domain adaptation for text classification via
  distancenet-bandits.
\newblock In \emph{Proceedings of the AAAI Conference on Artificial
  Intelligence}, volume~34, 7830--7838.

\bibitem[{Gupta, Singh, and Roth(2017)}]{gupta2017entity}
Gupta, N.; Singh, S.; and Roth, D. 2017.
\newblock Entity linking via joint encoding of types, descriptions, and
  context.
\newblock In \emph{Proceedings of the 2017 Conference on Empirical Methods in
  Natural Language Processing}, 2681--2690.

\bibitem[{He et~al.(2020)He, Fan, Wu, Xie, and Girshick}]{he2019momentum}
He, K.; Fan, H.; Wu, Y.; Xie, S.; and Girshick, R. 2020.
\newblock Momentum contrast for unsupervised visual representation learning.
\newblock In \emph{Proceedings of the IEEE/CVF Conference on Computer Vision
  and Pattern Recognition}, 9729--9738.

\bibitem[{Hou et~al.(2020)Hou, Wang, He, and Zhou}]{hou2020improving}
Hou, F.; Wang, R.; He, J.; and Zhou, Y. 2020.
\newblock Improving Entity Linking through Semantic Reinforced Entity
  Embeddings.
\newblock In \emph{Proceedings of the 58th Annual Meeting of the Association
  for Computational Linguistics}, 6843--6848.

\bibitem[{Hu et~al.(2020)Hu, Dong, Wang, Chang, and Sun}]{hu2020gpt}
Hu, Z.; Dong, Y.; Wang, K.; Chang, K.-W.; and Sun, Y. 2020.
\newblock Gpt-gnn: Generative pre-training of graph neural networks.
\newblock In \emph{Proceedings of the 26th ACM SIGKDD International Conference
  on Knowledge Discovery \& Data Mining}, 1857--1867.

\bibitem[{Klie, de~Castilho, and Gurevych(2020)}]{klie2020zero}
Klie, J.-C.; de~Castilho, R.~E.; and Gurevych, I. 2020.
\newblock From zero to hero: Human-in-the-loop entity linking in low resource
  domains.
\newblock In \emph{Proceedings of the 58th Annual Meeting of the Association
  for Computational Linguistics}, 6982--6993.

\bibitem[{Kolitsas and Ganea(2018)}]{Kolitsas2018ACL}
Kolitsas, N.; and Ganea, O.-E. 2018.
\newblock End-to-End Neural Entity Linking.
\newblock In \emph{Proceedings of the 56th Annual Meeting of the Association
  for Computational Linguistics}.

\bibitem[{Lample et~al.(2018)Lample, Conneau, Ranzato, Denoyer, and
  J{\'e}gou}]{lample2018word}
Lample, G.; Conneau, A.; Ranzato, M.; Denoyer, L.; and J{\'e}gou, H. 2018.
\newblock Word translation without parallel data.
\newblock In \emph{International Conference on Learning Representations}.

\bibitem[{Le and Titov(2018)}]{le2018improving}
Le, P.; and Titov, I. 2018.
\newblock Improving entity linking by modeling latent relations between
  mentions.
\newblock In \emph{Proceedings of the 56th Annual Meeting of the Association
  for Computational Linguistics}.

\bibitem[{Li et~al.(2013)Li, Liang, Ding, Yang, and Pan}]{li2013feature}
Li, J.; Liang, X.; Ding, W.; Yang, W.; and Pan, R. 2013.
\newblock Feature engineering and tree modeling for author-paper identification
  challenge.
\newblock In \emph{KDD Cup 2013 Workshop}, 5.

\bibitem[{Liu, Qiu, and Huang(2017)}]{liu2017adversarial}
Liu, P.; Qiu, X.; and Huang, X. 2017.
\newblock Adversarial multi-task learning for text classification.
\newblock In \emph{Proceedings of the 56th Annual Meeting of the Association
  for Computational Linguistics}, 1--10.

\bibitem[{Logeswaran et~al.(2019)Logeswaran, Chang, Lee, Toutanova, Devlin, and
  Lee}]{logeswaran2019zero}
Logeswaran, L.; Chang, M.-W.; Lee, K.; Toutanova, K.; Devlin, J.; and Lee, H.
  2019.
\newblock Zero-Shot Entity Linking by Reading Entity Descriptions.
\newblock In \emph{Proceedings of the 57th Annual Meeting of the Association
  for Computational Linguistics}.

\bibitem[{Louppe et~al.(2016)Louppe, Al-Natsheh, Susik, and
  Maguire}]{louppe2016ethnicity}
Louppe, G.; Al-Natsheh, H.~T.; Susik, M.; and Maguire, E.~J. 2016.
\newblock Ethnicity sensitive author disambiguation using semi-supervised
  learning.
\newblock In \emph{international conference on knowledge engineering and the
  semantic web}, 272--287. Springer.

\bibitem[{Qiao et~al.(2019)Qiao, Du, Fu, Wang, and Zhou}]{qiao2019unsupervised}
Qiao, Z.; Du, Y.; Fu, Y.; Wang, P.; and Zhou, Y. 2019.
\newblock Unsupervised author disambiguation using heterogeneous graph
  convolutional network embedding.
\newblock In \emph{2019 IEEE international conference on big data (Big Data)},
  910--919. IEEE.

\bibitem[{Qiu et~al.(2020)Qiu, Chen, Dong, Zhang, Yang, Ding, Wang, and
  Tang}]{qiu2020gcc}
Qiu, J.; Chen, Q.; Dong, Y.; Zhang, J.; Yang, H.; Ding, M.; Wang, K.; and Tang,
  J. 2020.
\newblock Gcc: Graph contrastive coding for graph neural network pre-training.
\newblock In \emph{Proceedings of the 26th ACM SIGKDD International Conference
  on Knowledge Discovery \& Data Mining}, 1150--1160.

\bibitem[{Roy et~al.(2013)Roy, De~Cock, Mandava, Savanna, Dalessandro, Perlich,
  Cukierski, and Hamner}]{roy2013microsoft}
Roy, S.~B.; De~Cock, M.; Mandava, V.; Savanna, S.; Dalessandro, B.; Perlich,
  C.; Cukierski, W.; and Hamner, B. 2013.
\newblock The microsoft academic search dataset and kdd cup 2013.
\newblock In \emph{KDD cup 2013 workshop}, 1--6.

\bibitem[{Shi et~al.(2018)Shi, Feng, Huang, Zhang, Ji, Liao, and
  Huang}]{shi2018genre}
Shi, G.; Feng, C.; Huang, L.; Zhang, B.; Ji, H.; Liao, L.; and Huang, H.-Y.
  2018.
\newblock Genre separation network with adversarial training for cross-genre
  relation extraction.
\newblock In \emph{Proceedings of the 2018 Conference on Empirical Methods in
  Natural Language Processing}, 1018--1023.

\bibitem[{Tang et~al.(2008)Tang, Zhang, Yao, Li, Zhang, and
  Su}]{tang2008arnetminer}
Tang, J.; Zhang, J.; Yao, L.; Li, J.; Zhang, L.; and Su, Z. 2008.
\newblock Arnetminer: extraction and mining of academic social networks.
\newblock In \emph{Proceedings of the 14th ACM SIGKDD international conference
  on Knowledge discovery and data mining}, 990--998.

\bibitem[{Van~der Maaten and Hinton(2008)}]{maaten2008visualizing}
Van~der Maaten, L.; and Hinton, G. 2008.
\newblock Visualizing data using t-SNE.
\newblock \emph{Journal of machine learning research}, 9(11).

\bibitem[{Wang et~al.(2020)Wang, Wan, Wen, Li, Jia, Zhang, and
  Wang}]{wang2020author}
Wang, H.; Wan, R.; Wen, C.; Li, S.; Jia, Y.; Zhang, W.; and Wang, X. 2020.
\newblock Author name disambiguation on heterogeneous information network with
  adversarial representation learning.
\newblock In \emph{Proceedings of the AAAI Conference on Artificial
  Intelligence}, volume~34, 238--245.

\bibitem[{Wolf et~al.(2020)Wolf, Debut, Sanh, Chaumond, Delangue, Moi, Cistac,
  Rault, Louf, Funtowicz, Davison, Shleifer, von Platen, Ma, Jernite, Plu, Xu,
  Scao, Gugger, Drame, Lhoest, and Rush}]{wolf-etal-2020-transformers}
Wolf, T.; Debut, L.; Sanh, V.; Chaumond, J.; Delangue, C.; Moi, A.; Cistac, P.;
  Rault, T.; Louf, R.; Funtowicz, M.; Davison, J.; Shleifer, S.; von Platen,
  P.; Ma, C.; Jernite, Y.; Plu, J.; Xu, C.; Scao, T.~L.; Gugger, S.; Drame, M.;
  Lhoest, Q.; and Rush, A.~M. 2020.
\newblock Transformers: State-of-the-Art Natural Language Processing.
\newblock In \emph{Proceedings of the 2020 Conference on Empirical Methods in
  Natural Language Processing: System Demonstrations}, 38--45.

\bibitem[{Wu et~al.(2018)Wu, Xiong, Yu, and Lin}]{wu2018unsupervised}
Wu, Z.; Xiong, Y.; Yu, S.~X.; and Lin, D. 2018.
\newblock Unsupervised feature learning via non-parametric instance
  discrimination.
\newblock In \emph{Proceedings of the IEEE Conference on Computer Vision and
  Pattern Recognition}, 3733--3742.

\bibitem[{Xiong et~al.(2017)Xiong, Dai, Callan, Liu, and Power}]{xiong2017end}
Xiong, C.; Dai, Z.; Callan, J.; Liu, Z.; and Power, R. 2017.
\newblock End-to-end neural ad-hoc ranking with kernel pooling.
\newblock In \emph{Proceedings of the 40th International ACM SIGIR Conference
  on Research and Development in Information Retrieval}, 55--64.

\bibitem[{Yamada et~al.(2020)Yamada, Asai, Shindo, Takeda, and
  Matsumoto}]{yamada2020luke}
Yamada, I.; Asai, A.; Shindo, H.; Takeda, H.; and Matsumoto, Y. 2020.
\newblock LUKE: Deep Contextualized Entity Representations with Entity-aware
  Self-attention.
\newblock In \emph{Proceedings of the 2020 Conference on Empirical Methods in
  Natural Language Processing (EMNLP)}, 6442--6454.

\bibitem[{Yang et~al.(2019)Yang, Dai, Yang, Carbonell, Salakhutdinov, and
  Le}]{yang2019xlnet}
Yang, Z.; Dai, Z.; Yang, Y.; Carbonell, J.; Salakhutdinov, R.~R.; and Le, Q.~V.
  2019.
\newblock Xlnet: Generalized autoregressive pretraining for language
  understanding.
\newblock \emph{Advances in neural information processing systems}, 32.

\bibitem[{You et~al.(2020)You, Chen, Sui, Chen, Wang, and Shen}]{you2020graph}
You, Y.; Chen, T.; Sui, Y.; Chen, T.; Wang, Z.; and Shen, Y. 2020.
\newblock Graph Contrastive Learning with Augmentations.
\newblock \emph{Advances in Neural Information Processing Systems}, 33.

\bibitem[{Zemlyanskiy et~al.(2021)Zemlyanskiy, Gandhe, He, Kanagal, Ravula,
  Gottweis, Sha, and Eckstein}]{zemlyanskiy2021docent}
Zemlyanskiy, Y.; Gandhe, S.; He, R.; Kanagal, B.; Ravula, A.; Gottweis, J.;
  Sha, F.; and Eckstein, I. 2021.
\newblock DOCENT: Learning Self-Supervised Entity Representations from Large
  Document Collections.
\newblock \emph{arXiv preprint arXiv:2102.13247}.

\bibitem[{Zhai et~al.(2020)Zhai, Lu, Ye, Shan, Chen, Ji, and Tian}]{zhai2020ad}
Zhai, Y.; Lu, S.; Ye, Q.; Shan, X.; Chen, J.; Ji, R.; and Tian, Y. 2020.
\newblock Ad-cluster: Augmented discriminative clustering for domain adaptive
  person re-identification.
\newblock In \emph{Proceedings of the IEEE/CVF Conference on Computer Vision
  and Pattern Recognition}, 9021--9030.

\bibitem[{Zhang and Al~Hasan(2017)}]{zhang2017name}
Zhang, B.; and Al~Hasan, M. 2017.
\newblock Name disambiguation in anonymized graphs using network embedding.
\newblock In \emph{Proceedings of the 2017 ACM on Conference on Information and
  Knowledge Management}, 1239--1248.

\bibitem[{Zhang et~al.(2018{\natexlab{a}})Zhang, Huang, Yu, Zhang, and
  Chawla}]{zhang2018camel}
Zhang, C.; Huang, C.; Yu, L.; Zhang, X.; and Chawla, N.~V. 2018{\natexlab{a}}.
\newblock Camel: Content-Aware and Meta-path Augmented Metric Learning for
  Author Identification.
\newblock In \emph{Proceedings of the 2018 World Wide Web Conference on World
  Wide Web}, 709--718.

\bibitem[{Zhang et~al.(2015)Zhang, Tang, Yang, Pei, and Yu}]{zhang2015cosnet}
Zhang, Y.; Tang, J.; Yang, Z.; Pei, J.; and Yu, P.~S. 2015.
\newblock Cosnet: Connecting heterogeneous social networks with local and
  global consistency.
\newblock In \emph{Proceedings of the 21th ACM SIGKDD International Conference
  on Knowledge Discovery and Data Mining}, 1485--1494.

\bibitem[{Zhang et~al.(2018{\natexlab{b}})Zhang, Zhang, Yao, and
  Tang}]{zhang2018name}
Zhang, Y.; Zhang, F.; Yao, P.; and Tang, J. 2018{\natexlab{b}}.
\newblock Name Disambiguation in AMiner: Clustering, Maintenance, and Human in
  the Loop.
\newblock In \emph{Proceedings of the 24th ACM SIGKDD International Conference
  on Knowledge Discovery \& Data Mining}, 1002--1011.

\end{thebibliography}
	\clearpage
	\section{Appendix}

\hide{
\begin{table}[htbp]
	\caption{Hyper-parameters of \model.}
	\label{tbl:hyper-parameters}
	\small
	\begin{tabular}{l |l}
		\toprule
		Batch size             & 32   \\
		Margin				   & 1.0 \\
		Gradient descent		& Adam \\
		Expert encoder learning rate & 2e-5\\
		Metric learning rate	& 2e-3\\
		Max epoch		& 10 \\
		Learning rate decay		   & Exponential ($\gamma$ = 0.96)\\
		BERT MLP			& $\mathbb{R}^{768\times300}$\\
		Metric kernels		& 21 \\
		Metric MLP 			& $\mathbb{R}^{21\times10}$, $\mathbb{R}^{10\times1}$\\
		\bottomrule
	\end{tabular}
	\normalsize
\end{table}
}
\begin{table}[htbp]
\newcolumntype{?}{!{\vrule width 1pt}}
	
	\caption{Hyper-parameters of \model.}
	\centering 
	\label{tbl:hyper}
	\small
	\begin{tabular}{@{}l?l@{}}
		\toprule

		\multicolumn{2}{c}{\textbf{Model architecture} }  \\
		\midrule
		BERT MLP			& $\mathbb{R}^{768\times512}$\\
		\#Kernels $K$		& 21 \\
		Metric MLP 			& $\mathbb{R}^{21\times21}$, $\mathbb{R}^{21\times1}$\\
		Margin $m$				   & 1.0 \\
		Adversarial MLP  &$\mathbb{R}^{512\times100}$, $\mathbb{R}^{100\times2}$\\
		LeakyReLU              & Negative slope is 0.2 \\

		\midrule
		\multicolumn{2}{c}{\textbf{Learning rate} }  \\
		\midrule
		$\mu_g$ for encoder& 2e-5\\
		$\mu_f$ for metric function & 2e-3\\
		$\mu_h$ for discriminator& 1e-3\\ 
		Learning rate decay		   & Exponential decay = 0.96 \\

		\midrule
		\multicolumn{2}{c}{\textbf{Loss function weight} }  \\
		\midrule
		$\alpha$ for adversarial loss & 0.1\\
		$\beta$ for difference loss   & 0.1 \\
		$\gamma$ for external task loss   & 0.1 \\

		\midrule
		\multicolumn{2}{c}{\textbf{Batch size} }  \\
		\midrule
		For pre-training on AMiner data   & 32   \\
		For fine-tuning on External sources 		& 256	\\

		\midrule
		\multicolumn{2}{c}{\textbf{Others} }  \\
		\midrule
		\#Negative Instances & 9 \\
		\#Papers & 6 \\
		Optimization algorithm		& Adam \\
		Pre-training epochs		& 20 \\
		Fine-tuning epochs      & 1 \\

		\bottomrule
	\end{tabular}
	\normalsize
\end{table}

\subsection{Implementation Details}
The detailed hyper-parameters are listed in Table~\ref{tbl:hyper}.

\subsubsection{Running Environment}
We implement \smodel by Python 3.6.8, PyTorch 1.1.0, and conduct the experiments on an Enterprise Linux Server with 56 Intel(R) Xeon(R) Gold 5120 CPU @ 2.20GHz, and a single NVIDIA Tesla V100 SXM2 with 32GB memory size.

\subsubsection{Pre-training Module}

The dimension of the BERT output embedding is 768, which is then fed into a one-layer MLP in Eq.\eqref{eq:encoder} to get the corresponding outputs:

\beq{
	\text{MLP}(\textbf{X}) = \text{Tanh}(\textbf{W}^T\textbf{X}),
}

\noindent where $\textbf{W} \in \mathbb{R}^{768\times 512}$. 

The two-layers MLP on top of the metric function in Eq.\eqref{eq:interaction_metric_function} is defined as follows:

\beq{
  \text{MLP}(\textbf{X}) =  \tanh (\textbf{W}_2^T\text{LeakyReLU} (\textbf{W}_1^T\textbf{X})), 
}

\noindent where $\textbf{W}_1 \in \mathbb{R}^{21 \times 21}$ and $\textbf{W}_2 \in \mathbb{R}^{21\times 1}$.

In Eq.\eqref{eq:kernel}, we use 21 kernels, where $\mu$ is from 0 to 1 with interval 0.05.  The kernel with $\mu=0.0$ and $\sigma=10^{-3}$ corresponds to the exact matching kernel, while $\sigma$  is set as 0.1 for other kernels capture the semantic matches at different scales. 
The margin $m$ of the triplet loss in Eq.\eqref{eq:triplet_loss} is set to 1.0.

\subsubsection{Adversarial Fine-tuning Module}

The two-layers MLP for the domain discriminator in Eq.\eqref{eq:adversarial} is defined as:

\beq{
  \text{MLP}(\textbf{X}) =  \textbf{W}_2^T\text{LeakyReLU} (\textbf{W}_1^T\textbf{X}), 
}

\noindent where $\textbf{W}_1^T \in \mathbb{R}^{512\times100}$ and $\textbf{W}_2^T \in \mathbb{R}^{100\times2}$. Likewise, the classifier $\tilde{h}$ of the Task Predictor is defined the same as above.



\begin{table}[t]
		{\caption{\textbf{Features extracted for GBDT model}. \small{$p$: paper, $a$: an author in $p$, $c$: a candidate expert whose name is similar as $a$.} \label{tb:features}}} 
		\vspace{-0.08in}
		\centering 
	\footnotesize
		{\renewcommand{\arraystretch}{1}%
			{
				\setlength{\extrarowheight}{1pt}
				\begin{tabular}{
						@{}c@{ } l@{}}
					\noalign{ \hrule height 1pt}
					\textbf{No.}   & \textbf{Feature description} \\ \hline
					\textbf{1}      &  The number of the papers of $c$\\ 
					\hdashline
					\textbf{2}      &  The number of the coauthors of $a$ in $p$\\  
					\textbf{3}      &  The number of the coauthors of $c$\\  
					\textbf{4}      &  The number of the same coauthors between $a$ and $c$\\  
					\textbf{5}      &  Ratio of the same coauthors between $a$ and $c$ in $p$'s coauthor names\\ 
					\textbf{6}      &  Ratio of the same coauthors  between $a$ and $c$  in $c$'s coauthor names\\ \hdashline		
					\textbf{7}      &  Frequency of $a$'s affiliation in $c$'s affiliations\\ 
					\textbf{8}      &  Ratio of $a$'s affiliation in $c$'s affiliations\\ 
					\textbf{9}      &  Cosine similarity between $a$'s affiliation and $c$'s affiliations\\ 
					\textbf{10}    &  Jaccards similarity between $a$'s affiliation and $c$'s affiliations\\  \hdashline
					\textbf{11}    &  Distinct number of venues of $c$\\
					\textbf{12}    &  Frequency of $p$'s venue in $c$\\ 
					\textbf{13}    &  Ratio of $p$'s venue in $c$\\ 
					\textbf{14}    &  Cosine similarity between $p$'s venue and $c$'s venues \\ 
					\textbf{15}    &  Jaccards similarity between $p$'s venue and $c$'s venues \\  \hdashline
					\textbf{16}    &  Cosine similarity between $p$'s title and $c$'s titles\\ 
					\textbf{17}    &  Jaccards similarity between $p$'s title and $c$'s titles\\  \hdashline
					\textbf{18}    &  Distinct number of keywords in $c$\\ 	
					\textbf{19}    &  Frequency of $p$'s keywords of $c$\\ 			
					\textbf{20}    &  Ratio of $p$'s keywords in $c$\\ 	 
					\textbf{21}    &  Cosine similarity between $p$'s keywords and $c$'s keywords\\ 
					\textbf{22}    &  Jaccards similarity between $p$'s keywords and $c$'s keywords\\   
					
					\noalign{\hrule height 1pt}
			\end{tabular}}
			
		}
\end{table}

\subsection{Training and Test Settings}

\subsubsection{Evaluation of the Pre-training Module.} Pre-training is performed only on the AMiner dataset.  

\vpara{Training Settings.} 
For the AMiner dataset shown in Table~\ref{tb:dataset}, we first filter the experts with less than 6 papers to satisfy the best number of the sampled papers for each expert.
For each expert in the remaining AMiner dataset, we randomly sample $L=6$ papers to comprise an expert instance. For each anchor expert instance, we sample 1 positive counterpart together with 9 negative counterparts, which results in 4,800 positive instance pairs and 43,200 negative instance pairs in total.
For embedding a paper, we use [CLS] + title + keywords + name + organization + venue + [SEP] as the BERT input. The maximal length of the input tokens is set as 208.

\vpara{Test Settings.} Author Identification and Paper Clustering are two intrinsic tasks for testing the pre-training module.

\begin{itemize}[leftmargin=*]
	\item 
	\textbf{Author Identification.}
	follows the second task of the name disambiguation competition$^{10}$. Thus, we adopt the same test set as the competition. 
	We randomly sample 17 negative experts together the ground truth expert to comprise the candidate list for a queried paper. The maximal number of papers sampled for each candidate expert is set as 100. For testing, we first leverage the pre-trained expert encoder $g$ to generate the paper embedding for each candidate expert, then use the pre-trained metric function $f$ to measure the similarity between a paper and each candidate expert. Finally, we rank all the candidate experts for a paper according to their similarity scores and return the top expert as the expert to be linked. 
	
	\item 
	\textbf{Paper Clustering.} follows the first task of the name disambiguation competition$^{11}$.
	We directly use its test set. For testing, we apply the pre-trained expert encoder $g$ to get the paper embeddings and then use the HAC algorithm to cluster the papers belonging to a same expert together. For a fair comparison, the true number of clusters in each name are provided to the HAC algorithm.
	
\end{itemize}

\vpara{Implementation of baselines.}
The heuristic features of GBDT~\cite{efimov2013kdd, li2013feature} are specified in Table~\ref{tb:features}. Apart from GBDT, We implement other baselines following their released code and settings. The training and test settings of all the baselines are the same as \pretrain.

\begin{itemize}[leftmargin=*]
	
	\item
	\textbf{Camel}~\cite{zhang2018camel}:
	
	\text{https://github.com/chuxuzhang/code\_Camel\_WWW2018}  
	
	\item
	\textbf{HetNetE}~\cite{chen2017task}:  
	
	\text{https://github.com/chentingpc/GuidedHeteEmbedding}

	\item
	\textbf{CONNA}~\cite{chen2020conna}:
	
	\text{https://github.com/BoChen-Daniel/TKDE-2019-CONNA} 

	\item
	\textbf{louppe at el}~\cite{louppe2016ethnicity}: 
	
	\text{https://github.com/glouppe/paper-author-disambiguation}
	
	\item
	\textbf{Zhang et al}~\cite{zhang2017name}: 
	
	\text{https://github.com/baichuan/disambiguation\_embedding}
	
	\item 
	\textbf{G/L-Emb}~\cite{zhang2018name}: 
	
	\text{https://github.com/neozhangthe1/disambiguation/}
	 
\end{itemize}

Both Camel and HetNetE define an additional loss function on the indirect relationships between a paper and an expert generated by the pre-defined meta-paths. 
We ignore this part for the following reasons. Since author identification in their work aim to predict the authors of an anonymous paper, name ambiguity is not the key problem to be tackled. Thus they can collect all the papers published in some related venues as the training data. The connectivity of the resultant heterogeneous graph is good enough to find the indirect relationships between any two nodes. However, to disambiguate experts with same names, we only collect the papers with the same author names. The resultant graph is too sparse to find the indirect relationships.

\subsubsection{Regarding the different baselines in the two tasks.} The two evaluation tasks pay attention to various objectives, where Author Identification matches papers with authors, but Paper Clustering solely matches papers. The distinguished goals require different baselines. Actually, we have tried the Paper Clustering baselines for Author Identification but obtained poor performance. The paper-to-paper pseudo label that under-performs the paper-to-expert pseudo label on Author Identification also indicates the representation learning is better to be close with the goal of downstream tasks. The proposed expert-to-expert pseudo label can represent experts as well as papers when contrasting experts based on the interactions of their papers, resulting in a good performance on both tasks.

\subsubsection{Evaluation of the Fine-tuning Module}
We evaluate of the fine-tuning module on two external datasets, News and LinkedIn, by the extrinsic task of external expert linking. We first fine-tune \spretrain on both AMiner and the unlabeled external dataset, and evaluate it on the manually-labeled test set.

\vpara{News.}
The fine-tuning settings for news are as follows:

\begin{itemize}[leftmargin=*]
	
\item
\textbf{Training.} 
We use 20,658 news articles for fine-tuning. We divide the contextual text of a name in a news article into sentences and treat each sentence as a piece of support information. We extract six sentences before and after a name in a news article as the support information. The maximal length of the input tokens for both the shared encoder and the private encoder is set as 64. 
	
\item
\textbf{Test.} 
We annotate 1,622 names in news articles with linkages to AMiner experts for testing. We choose the candidates by name variants and sample up to 100 papers for each candidate. 
\end{itemize}

\vpara{LinkedIn.}
The settings for LinkedIn are as follows:

\begin{itemize}[leftmargin=*]
	
\item
\textbf{Training.} 
We use 50,000 LinkedIn homepages for fine-tuning. We select three common semi-structured attributes including affiliation, skills and summary in a homepage as the support information. The affiliation and the concatenated keywords in skills are separate pieces of support information. The long-text summary is divided into multiple pieces of support information. The maximal length of the input tokens for both the shared encoder and the private encoder is set as 64.

\item
\textbf{Test.} 
We annotate 1,329 linkages between LinkedIn  users and  AMiner experts for testing. Other settings are the same as News.
\end{itemize}

\hide{
\subsubsection{Labels in the External Sources}
We study whether the model performance can be improved by adding additional labeled data from the external sources.

\vpara{News.} We sample 202 news-AMiner linkages as the additional labels on news data. For the ground truth expert of each name in news, we search all the candidate experts by name variants as the negative counterparts. The prediction loss on these external labels are the same as Eq.\eqref{eq:triplet_loss} based on the shared generator.

\vpara{LinkedIn.} We sample 366 Linkedin-AMiner linkages as the additional labels on LinkedIn data. Other settings are the same as above.
}

\end{document}